\documentclass[aps,prx,reprint,superscriptaddress,nofootinbib,longbibliography]{revtex4-2}

\usepackage{amsmath,amssymb, amsthm,amstext}
\usepackage{natbib}
\usepackage{graphicx}
\usepackage{color}
\usepackage{array, enumerate}
\usepackage{bm}
\usepackage{multirow}
\usepackage[breaklinks,colorlinks,citecolor=blue]{hyperref}
\usepackage{braket}
\usepackage{txfonts}
\usepackage{physics}
\usepackage[normalem]{ulem}
\usepackage{esint}

\def\be{\begin{equation}}
\def\ee{\end{equation}}

\begin{document}

\title{Self-lensing of electromagnetic counterparts to extreme mass ratio inspirals}

\author{Ruwen Zhou}
\email{zhouruwen@westlake.edu.cn}
\affiliation{Department of Astronomy, Westlake University, Hangzhou 310030, Zhejiang Province, China}
\affiliation{Tsung-Dao Lee Institute, Shanghai Jiao-Tong University, Shanghai, 520 Shengrong Road, 201210, People’s Republic of China}

\author{Zhen Pan}
\email{zhpan@sjtu.edu.cn}
\affiliation{Tsung-Dao Lee Institute, Shanghai Jiao-Tong University, Shanghai, 520 Shengrong Road, 201210, People’s Republic of China}
\affiliation{School of Physics \& Astronomy, Shanghai Jiao-Tong University, Shanghai, 800 Dongchuan Road, 200240, People’s Republic of China}

\begin{abstract}
Extreme mass ratio inspirals  on inclined orbits can produce recurrent electromagnetic flares by shocking gas at accretion-disk crossings. Using forward ray tracing in Kerr spacetime, we investigate how self-lensing by the central supermassive black hole (SMBH) reshapes these flares. Flares emitted behind the SMBH can be strongly magnified, whereas near-side flares remain almost unmagnified, producing strong-weak alternation even for identical intrinsic emission. As Lense-Thirring precession shifts the crossing azimuths, it generates long-term amplitude envelopes whose spacing can be used to infer the precession period. Lense-Thirring precession can also increase the cumulative geometric lensing probability over longer observing windows. Within individual flares, Keplerian motion of hotspots co-rotating with the disk produces time-dependent magnification, which can generate shoulders or double peaks and change the effective width. We also propose a spectral diagnostic to distinguish self-lensing from external lensing when their light curves are similar: self-lensing can amplify later, cooler emission and produce a softer secondary peak, whereas delayed images from a static external lens repeat the same intrinsic spectral evolution.
\end{abstract}

\date{\today}

\maketitle

\section{Introduction}

Extreme mass ratio inspirals (EMRIs) are one of the primary targets for future spaceborne gravitational wave (GW) missions, LISA~\citep{LISA}, Taiji~\citep{Taiji}, and Tianqin~\citep{TianQin}. However, the formation rates and relative contributions of different EMRI formation channels remain uncertain. In the classical dry (loss-cone) channel, stellar-mass black holes (sBHs) within the nuclear stellar cluster around a supermassive black hole (SMBH) can be gravitationally scattered onto low-angular-momentum orbits, where GW emission becomes prominent~\citep{Sigurdsson1997,Hopman_2005,Merritt2011}. Another dynamical pathway is the Hills mechanism, in which a compact binary is tidally disrupted by the SMBH, ejecting one component while capturing the other onto a bound orbit~\citep{Hills1988nature,Miller2005apjl,Raveh2021}. In contrast to these gas-independent dynamical channels, accretion disks in active galactic nuclei (AGNs) provide an additional, gas-assisted pathway. sBHs in the nuclear stellar cluster can be captured by the accretion disk and subsequently migrate inward under torques from density waves excited in the disk~\citep{PanYang2021}. This wet channel may be more efficient in producing EMRIs~\citep{PanYang2021,PanLyuYang2021}. The relative contributions of these formation channels are expected to be measured by future spaceborne GW missions~\citep{PanYang2021,PanLyuYang2021,SunLiPanYang_2026,CuiHanPan_2ndSpin_2025,Vazquez-Aceves2022}.

In addition to GW emission, EMRIs in gas-rich environments are expected to produce electromagnetic emission that may be detectable as quasi-periodic oscillations (QPOs) when the EMRI is embedded in the accretion disk~\citep{MillihertQPO}, or as quasi-periodic eruptions (QPEs) when its orbit is misaligned with the disk~\citep{ScienceopportunitieswetEMRI}. In recent years, more than ten X-ray QPE sources have been discovered in nearby galactic nuclei~\citep{GSN069discover2019_1stQPE,RXJ1301discover2013,RXJ1301discover2020_2ndQPE,eRO-QPE1-2discover2021,eRO-QPE3-4discover2024,eRO-QPE5discover2025,Anskydiscover2025,AT2022discover2025,J2344discover2026}. Though a number of models have been proposed to explain the origin of QPEs, the EMRI+disk model is one of the most favored scenarios~\citep{stardisc_collisions2010,stardisc_collisions_applyGSN069_spin_2021,EMRI+TDE_QPE2023,precesiondisk_starcollision2023,low_inc_orbit_collision2023,CongZhou2024a,CongZhou2024b,secularEvolutionQPE2025,Jiang:2025jbd,wrap_disk_QPE_BRD_2026,Zhang:2026haz}. Connecting the observed flare sequence to EMRI dynamics requires understanding how photon propagation modifies the emission generated at each disk crossing.

Gravitational lensing is a natural consequence of photon propagation in curved spacetime~\citep{stardisc_collisions2010,Lensing_by_KerrBH_2020}. For disk-crossing flares, emission from nearly behind the SMBH can be strongly magnified, whereas near-side emission is dominated by an almost unmagnified direct image. Since successive crossings occur on approximately opposite sides of the SMBH, this contrast can produce alternating strong and weak flares even when their intrinsic emission is identical. The alignment evolves over time, however, as Lense-Thirring precession shifts the crossing sites in azimuth~\citep{2002Celestial_mechanicskerr}, modulating the magnification over many successive events. On the shorter timescale of a single flare, motion of the emitting gas can further change the magnification as the intrinsic emission evolves. Self-lensing therefore connects the observed variability to both the orbital precession and the motion of the emitting region.

In this work, we combine the forward ray-tracing framework developed by \citet{Forward_RTracing_2025} for Kerr spacetime with a geodesic disk-crossing sequence and a phenomenological intrinsic flare profile~\citep{flaremodel_GRB2005,eROqpe1Arcodia2022}. By comparing hotspots held fixed during each flare with those co-rotating with the disk at the local Keplerian rate, we distinguish the effects of Lense-Thirring precession between events from those of source motion during emission. We show how the spacing of the amplitude envelopes can be used to infer the nodal-precession period and quantify how the same precession increases the cumulative geometric lensing probability. Within individual flares, this comparison reveals how source motion introduces additional structure and changes the effective width. We also propose a spectral diagnostic to distinguish self-lensing from external lensing when their light curves are similar, using an illustrative cooling prescription to compare the spectra at the flare peaks~\citep{overlapping_LightCurve_by_External_lensing2026}.

The paper is organized as follows. Section~\ref{sec:method} describes the crossing sequence, emission model, and ray-tracing method. We then present the static-hotspot light curves, magnification maps, and lensing probabilities in Section~\ref{sec:static}, before turning to the temporal and spectral signatures of Keplerian hotspots in Section~\ref{sec:rotating}. Section~\ref{sec:discuss} discusses the implications and limitations of our results and presents our conclusions. The ray-tracing formalism, angular magnification profiles, and evolution of successive Keplerian flares are presented in Appendices~\ref{app:kerr_raytracing}, \ref{appsubsec:magnification_envelopes}, and \ref{subsec:keplerian_evolution}, respectively. Throughout this work, we use geometrized units with $G=c=1$ and express lengths and times in units of the central black-hole mass $M$.

\begin{figure*}[htb]
\centering
\includegraphics[width=\textwidth]{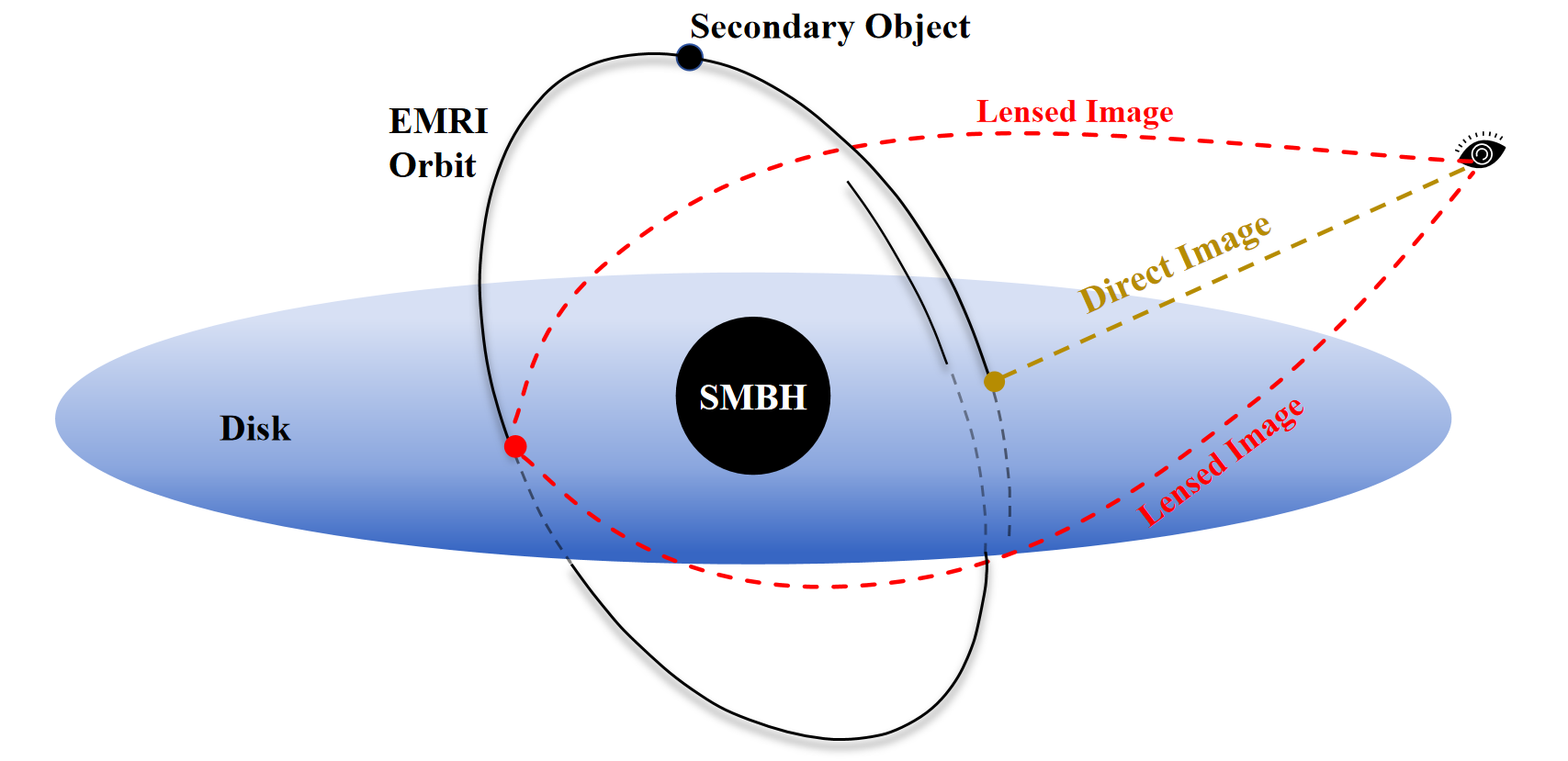}
\caption{ Schematic of the EMRI+disk collision geometry and the associated black-hole self-lensing configuration. Photons emitted at the collision sites marked by red and brown dots can follow multiple strongly deflected trajectories to a distant observer. The resulting images have different magnifications and propagation times, which modify the observed flare signal. }
\label{fig:schematic}
\end{figure*}

\section{Method}
\label{sec:method}

In the EMRI+disk scenario considered in this work, a compact object on an inclined orbit repeatedly crosses the equatorial accretion disk, producing a sequence of flares. Figure~\ref{fig:schematic} illustrates the geometry of this self-lensing configuration. Photons emitted during these disk-crossing events propagate through Kerr spacetime toward a distant observer. When the emitting region lies nearly behind the SMBH, strong gravitational deflection produces multiple images with different magnifications and arrival times. Our calculation combines the following three ingredients:
\begin{enumerate}
    \item a sequence of disk-crossing events generated by a bound timelike geodesic in Kerr spacetime;
    \item a phenomenological model for the intrinsic flare associated with each crossing;
    \item a general-relativistic ray-tracing calculation that integrates null geodesics in Kerr spacetime to map the intrinsic source emission into the lensed light curve received by a distant observer.
\end{enumerate}

The ray-tracing calculation follows the forward ray-tracing framework of \citet{Forward_RTracing_2025}. The underlying mathematical formalism is summarised in Appendix~\ref{app:kerr_raytracing}. Here, we focus on the ingredients required to construct the self-lensed light curves for both static and Keplerian hotspot models.

\subsection{Disk-Crossing Event Sequence and Lensing Geometry}
\label{subsec:qpe_event_sequence}

As illustrated in Figure~\ref{fig:schematic}, the secondary compact object is modeled as a test particle on a bound timelike geodesic around a Kerr black hole with spin parameter \(a\) ~\citep{2002Celestial_mechanicskerr}. The accretion disk is assumed to be geometrically thin and aligned with the equatorial plane. We label the chronologically ordered disk crossings by \(i=1,2,\dots\). The \(i\)-th crossing occurs at coordinate time \(t_i\), and its source spacetime coordinates are
\begin{equation}
    x_{s,i}^{\mu}
    =
    \left(
        t_i,\,
        r_{s,i},\,
        \theta_{s,i},\,
        \phi_{s,i}
    \right)
    =
    \left(
        t_i,\,
        r_{s,i},\,
        \frac{\pi}{2},\,
        \phi_{s,i}
    \right).
    \label{eq:crossing_source_position}
\end{equation}
We use the crossing time \(t_i\) as the phase reference for the phenomenological flare template, setting \(L_{\mathrm{int},i}(0)=L_{p,i}/e^3\).

We group the odd- and even-indexed crossings into two interleaved sequences, corresponding to the two disk-crossing branches. These branches are approximately separated by \(180^\circ\) in azimuth. We follow the long-term evolution of each branch by comparing events \(i\) and \(i+2\). In particular, Lense-Thirring precession causes the crossing azimuth \(\phi_{s,i}\) of each branch to drift, while apsidal precession produces smaller variations in the crossing radius \(r_{s,i}\).

For an equatorial source at an instantaneous position \((r_s,\theta_s,\phi_s)\) and a distant observer at \((r_o,\theta_o,\phi_o)\), we characterise the source-observer alignment by
\begin{equation}
    \phi_{o,\mathrm{rel}}
    \equiv
    \phi_o-\phi_s
    \pmod{2\pi}.
    \label{eq:observer_relative_azimuth}
\end{equation}
Here, \(r_o\) is taken to be sufficiently large that finite-distance effects are negligible. The configuration \(\phi_{o,\mathrm{rel}}\simeq0^\circ\) places the source on the near side of the black hole. By contrast, \(\phi_{o,\mathrm{rel}}\simeq180^\circ\) places it on the far side, close to the optical axis behind the SMBH. Kerr spacetime is axisymmetric, and the source is confined to the equatorial plane, \(\theta_s=\pi/2\). For fixed black-hole parameters and observer distance, \(r_s\), \(\theta_o\), and \(\phi_{o,\mathrm{rel}}\) therefore fully specify the instantaneous lensing geometry.

\subsection{Intrinsic Flare Model}
\label{subsec:intrinsic_flare}

We model the intrinsic emission associated with each disk crossing using a phenomenological fast-rise, exponential-decay profile. For the \(i\)-th event, we define the emission-time offset
\begin{equation}
    \tilde t
    \equiv
    t_e-t_i,
    \label{eq:local_emission_time}
\end{equation}
where \(t_e\) is the Boyer-Lindquist coordinate emission time and \(\tilde t=0\) is the disk-crossing reference time; \(\tilde t\) is not the emitter's proper time. We adopt the piecewise flare template of \citet{eROqpe1Arcodia2022}, which modifies the pulse prescription of \citet{flaremodel_GRB2005}:
\begin{equation}
L_{\mathrm{int},i}(\tilde t)=
\begin{cases}
0,
& \tilde t\leq t_p-t_{\mathrm{as}},
\\[4pt]
L_{p,i}
\exp\!\left[
\sqrt{\dfrac{\tau_1}{\tau_2}}
+
\dfrac{\tau_1}{t_p-t_{\mathrm{as}}-\tilde t}
\right],
& t_p-t_{\mathrm{as}}<\tilde t<t_p,
\\[10pt]
L_{p,i}
\exp\!\left[
-\dfrac{\tilde t-t_p}{\tau_2}
\right],
& \tilde t\geq t_p
\end{cases}
\label{eq:phenom_intrinsic_flare}
\end{equation}
Here, \(t_{\mathrm{as}}\equiv\sqrt{\tau_1\tau_2}\), and \(L_{p,i}\) is the peak luminosity. The parameter \(t_p\) is the interval from the disk crossing to the intrinsic peak. The parameters \(\tau_1\) and \(\tau_2\) control the rising and decaying branches, respectively. We set the luminosity at the disk crossing to \(L_{\mathrm{int},i}(0)=L_{p,i}/e^3\). This condition fixes \(\tau_1\) once \(t_p\) and \(\tau_2\) are chosen, with \(0<t_p<3\tau_2\), and implies \(t_p-t_{\mathrm{as}}<0\). Unless otherwise specified, we use the same \(t_p\) and \(\tau_2\) for all events and normalise their peak luminosities to \(L_{p,i}=1\). This isolates variations caused by the lensing geometry from intrinsic event-to-event luminosity variations.

\subsection{Source Motion and General-Relativistic Ray Tracing}
\label{subsec:forward_raytracing}

Because each flare has a finite duration, its emission location must be specified throughout the emitting phase. We denote the spacetime location of the emitting source associated with event \(i\) by
\begin{equation}
    x_{s,i}^{\mu}(\tilde t)
    =
    \left(
        t_i+\tilde t,\,
        r_{s,i},\,
        \frac{\pi}{2},\,
        \phi_{s,i}+\Omega_{s,i}\tilde t
    \right),
    \label{eq:source_trajectory_local_time}
\end{equation}
where
\begin{equation}
    \Omega_{s,i}
    =
    \begin{cases}
        0,
        & \text{static hotspot},\\[4pt]
        \Omega_{\mathrm K}(r_{s,i}),
        & \text{Keplerian hotspot},
    \end{cases}
    \quad
    \Omega_{\mathrm K}(r)
    =
    \frac{1}{r^{3/2}+a}.
    \label{eq:source_angular_velocity}
\end{equation}
At the disk crossing, both models pass through the same source position. In the static model, the spatial position of the emitting region remains fixed throughout the flare. In the Keplerian model, the source remains at \(r_{s,i}\) but co-rotates azimuthally with the equatorial disk. Accordingly, the relative azimuth defined in Equation~\eqref{eq:observer_relative_azimuth} remains constant during a static flare but evolves during a Keplerian flare.

At each emission-time offset \(\tilde t\), forward ray tracing integrates the null geodesics in Kerr spacetime that connect \(x_{s,i}^{\mu}(\tilde t)\) to the observer. Each solution defines an image \(\ell\) and its photon propagation time \(\tau_{i\ell}(\tilde t)\). The corresponding geometric area magnification \(\mu_{i\ell}(\tilde t)\), defined in the source-size convention of Appendix~\ref{appsubsec:magnification}, is used below as the flux weight of that image. The imaging conditions, magnification formalism, and image-order convention are summarized in Appendix~\ref{app:kerr_raytracing}. In the light curves discussed below, we retain the \(N=0\) and \(N=1\) images. Higher-order images are neglected because their magnifications are negligible for the source configurations considered here.

For each event and image, the emission time is mapped parametrically to the observer time according to
\begin{equation}
    t_{\mathrm{obs},i\ell}(\tilde t)
    =
    t_i+\tilde t+\tau_{i\ell}(\tilde t).
    \label{eq:image_observer_time_mapping}
\end{equation}
To isolate the geometric lensing response, we use the image magnifications as flux weights and neglect Doppler and gravitational frequency-shift transfer factors. Up to an overall distance normalisation, the corresponding image contribution is
\begin{equation}
    F_{i\ell}
    \!\left[t_{\mathrm{obs},i\ell}(\tilde t)\right]
    =
    \mu_{i\ell}(\tilde t)
    L_{\mathrm{int},i}(\tilde t).
    \label{eq:image_flux_parametric}
\end{equation}
The total observed light curve is obtained by summing all event and image contributions at the same observer time:
\begin{equation}
    F_{\mathrm{obs}}(t_{\mathrm{obs}})
    =
    \sum_i\sum_\ell F_{i\ell}(t_{\mathrm{obs}}).
    \label{eq:observed_lightcurve_general}
\end{equation}
Equations~\eqref{eq:image_observer_time_mapping} and \eqref{eq:image_flux_parametric} map the intrinsic emission to the observed light curve. The magnification sets the amplitude of each image contribution, whereas the propagation time sets its arrival time.

For a static hotspot, each image has a constant magnification and propagation time during an individual flare:
\begin{equation}
    \mu_{i\ell}^{\mathrm{stat}}
    \equiv
    \mu_{i\ell}(0),
    \qquad
    \tau_{i\ell}^{\mathrm{stat}}
    \equiv
    \tau_{i\ell}(0).
    \label{eq:static_transfer_quantities}
\end{equation}
Equation~\eqref{eq:observed_lightcurve_general} then reduces to
\begin{equation}
    F_{\mathrm{obs}}^{\mathrm{stat}}(t_{\mathrm{obs}})
    =
    \sum_i\sum_\ell
    \mu_{i\ell}^{\mathrm{stat}}
    L_{\mathrm{int},i}
    \!\left(
        t_{\mathrm{obs}}-t_i-\tau_{i\ell}^{\mathrm{stat}}
    \right).
    \label{eq:observed_lightcurve_static_local}
\end{equation}
Thus, each static image is a shifted and magnified copy of the intrinsic profile.

For a Keplerian hotspot, both the magnification and photon propagation time change as the source moves. The observed flare profile depends on when the strongest magnification occurs relative to the intrinsic luminosity peak. Magnification before, near, or after this peak mainly enhances the rising part, peak, or decaying part of the flare, respectively. Changes in photon propagation time also affect when these contributions reach the observer. As Lense-Thirring precession shifts the crossing azimuth between successive events on the same branch, the strongest magnification occurs at different stages of the intrinsic flare.

\begin{figure*}[htb]
\centering
\includegraphics[width=\textwidth]{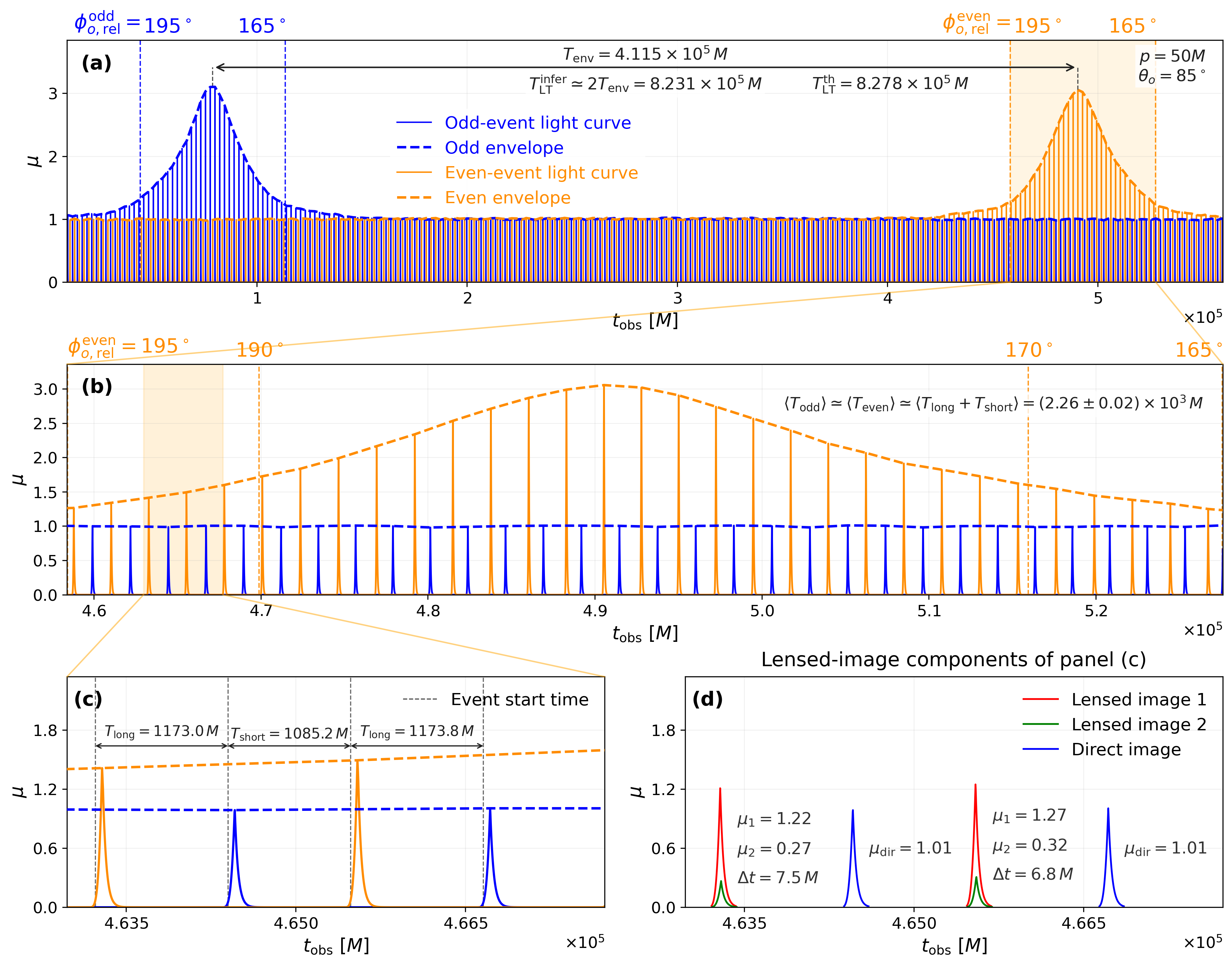}
\caption{ Self-lensing signatures in the observed light curve for the static-hotspot case with $p=50\,M$ and $\theta_o=85^\circ$. \textbf{Panel (a)} shows the long-timescale light curve decomposed into odd- and even-event branches, together with their lensing envelopes. As the observer-relative azimuth $\phi_{o,\mathrm{rel}}$ drifts due to Lense-Thirring precession, each branch is periodically brought close to the optical axis and forms a broad magnification envelope. Adjacent envelope maxima in the combined odd-even sequence are separated by \(T_{\mathrm{env}}=4.115\times10^5\,M\). The inferred full nodal-precession period is therefore \(T_{\mathrm{LT}}^{\mathrm{infer}} =2T_{\mathrm{env}}=8.231\times10^5\,M\), consistent with \(T_{\mathrm{LT}}^{\mathrm{th}}=8.278\times10^5\,M\). \textbf{Panel (b)} zooms into the even-event envelope over $\phi_{o,\mathrm{rel}}=195^\circ$ to $165^\circ$. The recurrence time within each branch remains approximately constant, as observed in GSN 069~\citep{GSN069discover2019_1stQPE,CongZhou2024a,CongZhou2024b}. \textbf{Panel (c)} shows the local short-timescale light curve near the envelope shoulder. Self-lensing naturally imprints a strong-weak flare pattern on top of the intrinsic long-short recurrence pattern. \textbf{Panel (d)} decomposes the flares in Panel (c) into low-order image components. We retain the $N=0$ and $N=1$ images because higher-order images are strongly demagnified. Far-side events produce components with different magnifications, $\mu_1$ and $\mu_2$, and a finite delay $\Delta t$. The direct image, with $\mu_{\mathrm{dir}}\simeq1$, traces the near-side emission pattern. }
\label{fig:Static_LightCurve}
\end{figure*}

\section{Self-Lensing of Static Hotspots}
\label{sec:static}

Here, we use the static hotspot prescription introduced in Section~\ref{subsec:forward_raytracing} to establish the basic self-lensing signatures of EMRI+disk flares. The emitting region remains fixed at the disk-crossing location throughout each flare. Equation~\eqref{eq:observed_lightcurve_static_local} therefore assigns every image of event \(i\) a constant magnification \(\mu_{i\ell}^{\mathrm{stat}}\) and propagation time \(\tau_{i\ell}^{\mathrm{stat}}\).

\subsection{Parameter Setup}
We adopt a representative parameter set for the examples below. The central SMBH has dimensionless spin $a=0.5$, and the secondary is a test particle following a timelike Kerr geodesic. The orbit has semi-latus rectum $p=50\,M$, eccentricity $e=0.1$, and inclination parameter $z_1=1/\sqrt{2}$, where $z_1=\cos\theta_{\min}$~\citep{2002Celestial_mechanicskerr}. We set the initial orbital phases to $\chi_{r,0}=0$, $\chi_{\theta,0}=90^\circ$, and $\phi_0=0$. We adopt $t_p=60\,M$, $\tau_2=30\,M$, and $L_{p,i}=1$ for all flares.

We place the observer at $r_o=10^4\,M$, which effectively represents an observer at infinity, and set $\theta_o=85^\circ$ and $\phi_o=210^\circ$. We track the two interleaved crossing branches by separating the ordered events into odd and even sequences. Adjacent events, $i$ and $i+1$, occur at approximately opposite azimuths and therefore sample near- and far-side geometries. Lense-Thirring precession causes both branch azimuths to drift with time~\citep{2002Celestial_mechanicskerr}.

We characterise each alignment using the observer-relative azimuth defined in Equation~\eqref{eq:observer_relative_azimuth}. Because every crossing lies in the equatorial plane, $\theta_s=90^\circ$, the observer polar angle $\theta_o$ specifies the polar geometry. Near $\phi_{o,\mathrm{rel}}=0^\circ$, the source lies on the near side and the direct image generally dominates. Near $\phi_{o,\mathrm{rel}}=180^\circ$, the source lies behind the black hole and strongly deflected images can be substantially magnified. We denote the two branch azimuths by $\phi_{o,\mathrm{rel}}^{\mathrm{odd}}$ and $\phi_{o,\mathrm{rel}}^{\mathrm{even}}$.

\begin{figure*}[t]
\centering
\includegraphics[width=\textwidth]{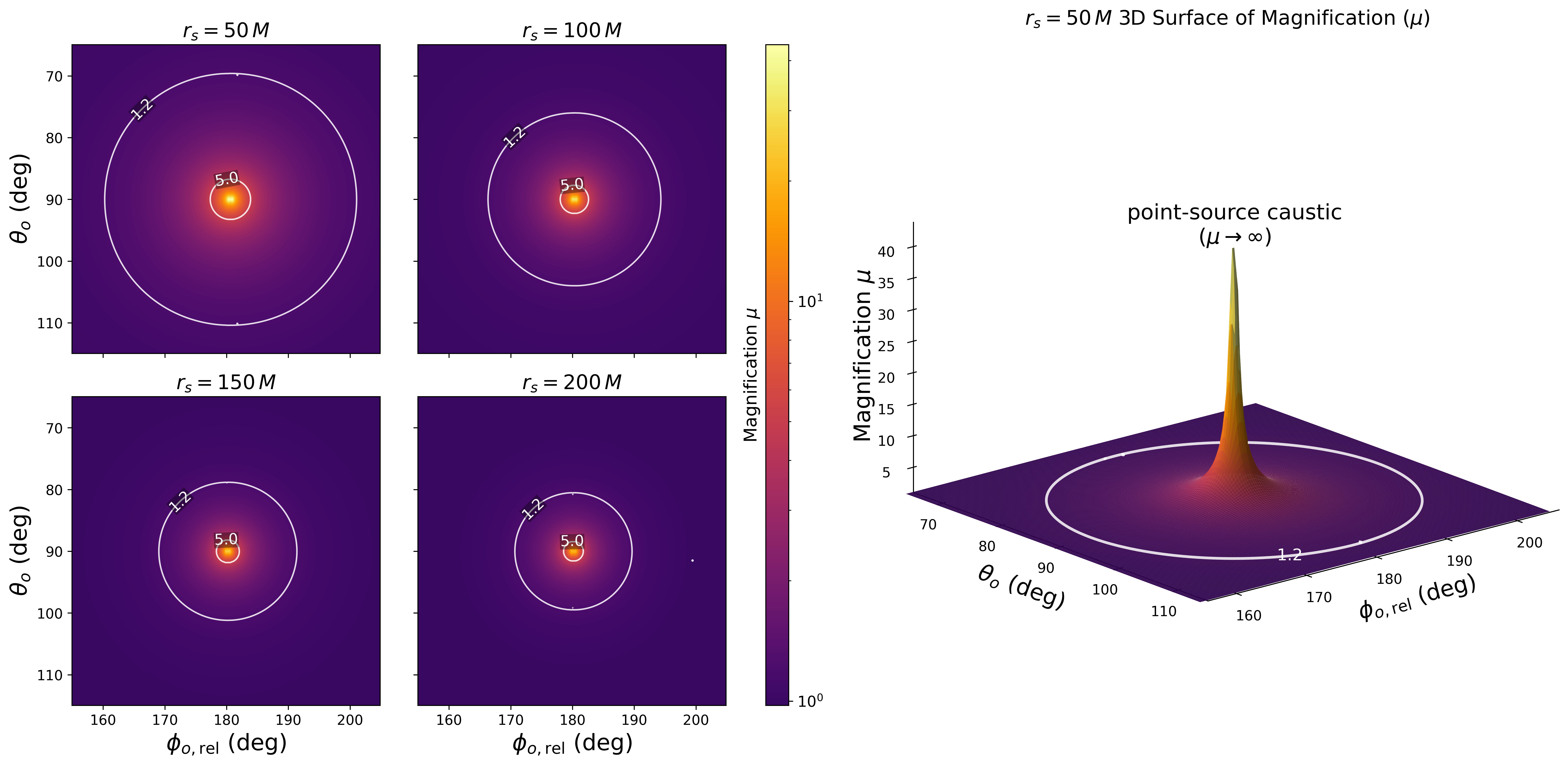}
\caption{ Self-lensing magnification maps on the observer sky. All flare events lie in the equatorial plane, $\theta_s=90^\circ$. Each pixel represents an observer-SMBH-source configuration specified by $(\theta_o,\phi_{o,\mathrm{rel}})$. The four left panels show two-dimensional maps for source orbital radii $r_s=50,100,150,$ and $200\,M$, over $65^\circ\leq\theta_o\leq115^\circ$ and $155^\circ\leq\phi_{o,\mathrm{rel}}\leq205^\circ$. White contours mark $\mu_{\mathrm{tot}}=1.2$ and $\mu_{\mathrm{tot}}=5$. The magnification rises sharply toward the nearly aligned configuration, $\theta_o\simeq90^\circ$ and $\phi_{o,\mathrm{rel}}\simeq180^\circ$. For random viewing directions, dividing the solid angle enclosed by a fixed-magnification contour by \(4\pi\) gives the geometric self-lensing probability for a fixed source position. The right panel shows the corresponding three-dimensional surface for $r_s=50\,M$. The central spike marks the point-source caustic limit, where the total magnification formally diverges, $\mu_{\mathrm{tot}}\rightarrow\infty$. A finite flare region would regularise this divergence through source averaging, which is not applied here. The finite peak displayed in the surface plot is therefore a numerical truncation of the point-source divergence, not a physical finite-source magnification. }
\label{fig:lensing_possibility_maps}
\end{figure*}
\begin{figure}[t]
    \centering
    \includegraphics[width=\columnwidth]{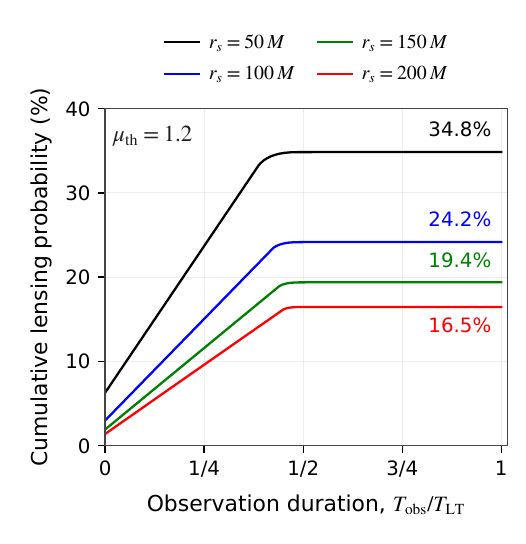}
    \caption{ Cumulative geometric lensing probability for \(r_s=50,100,150,\) and \(200\,M\), with \(\mu_{\mathrm{th}}=1.2\). The probability is the fraction of viewing directions from which at least one sampled flare exceeds the magnification threshold. The two crossing branches are approximated as antipodal at a common fixed radius, with uniform Lense-Thirring precession and dense phase sampling. As the observing window increases, the swept regions overlap and the probability approaches the full accessible belt fraction \(P_{\mathrm{belt}}\), whose values are annotated. The belt is covered by \(T_{\mathrm{obs}}=T_{\mathrm{LT}}/2\). The intercept is \(2P_{1.2}(r_s)\), because the calculation includes both approximately antipodal disk-crossing branches. }
    \label{fig:lt_lensing_probability}
\end{figure}
\begin{figure}[htb]
\centering
\includegraphics[width=\linewidth]{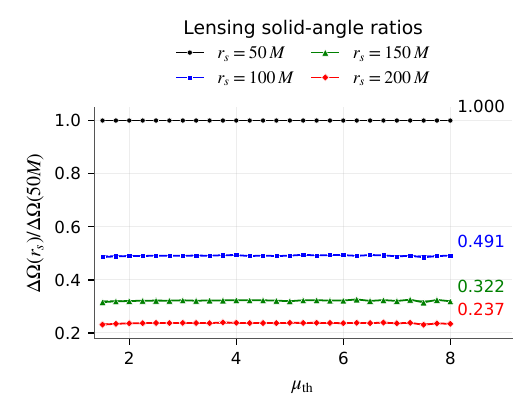}
\caption{ Self-lensing observer-sky solid-angle ratios as a function of magnification threshold, measured from Figure~\ref{fig:lensing_possibility_maps}. For each threshold \(\mu_{\mathrm{th}}\), \(\Delta\Omega_{\mu_{\mathrm{th}}}(r_s)\) is the solid angle over which \(\mu_{\mathrm{tot}}>\mu_{\mathrm{th}}\). The curves show the ratios for \(r_s=50,100,150,\) and \(200\,M\), normalized to the \(r_s=50\,M\) case. The ratios depend only weakly on the threshold. The annotated values are averages over the sampled thresholds and indicate an approximately inverse-radius dependence. }
\label{fig:lensing_possibility_scaling}
\end{figure}
\subsection{Self-Lensed Light Curve of Static Hotspots}
\label{subsec:static_lightcurve}
Figure~\ref{fig:Static_LightCurve}(a) shows 487 successive flares in the fiducial static-hotspot model. Lense-Thirring precession sweeps the odd- and even-event branches through the far-side alignment, producing broad magnification envelopes despite identical intrinsic flare profiles. Within each branch, the amplification peaks near \(\phi_{o,\mathrm{rel}}=180^\circ\) and decreases as the crossing site moves away from the optical axis. The two branches exhibit similar envelopes at different precessional phases, reflecting their approximately \(180^\circ\) azimuthal separation.

The recurrence of the envelope maxima provides a timing diagnostic of the underlying Lense-Thirring precession. From Panel~(a), the separation between adjacent envelope peaks in the combined odd-even sequence is $T_{\mathrm{env}}=4.115\times10^5\,M$. Because the two disk-crossing branches are separated by approximately $180^\circ$, the far-side lensing configuration is encountered twice during one full nodal-precession cycle. The precession period inferred from the envelope spacing is therefore
\begin{equation}
    T_{\mathrm{LT}}^{\mathrm{infer}}=2T_{\mathrm{env}}
    =8.231\times10^5\,M .
\end{equation}
For the same Kerr geodesic, the theoretical nodal-precession frequency is given by the difference between the azimuthal and polar fundamental frequencies~\citep{2002Celestial_mechanicskerr},
\begin{equation}
    \Omega_{\mathrm{LT}}=\Omega_\phi-\Omega_\theta .
\end{equation}
For a generic eccentric and inclined Kerr orbit, the exact expressions for \(\Omega_\phi\) and \(\Omega_\theta\) involve elliptic integrals, so we evaluate them numerically for the geodesic specified above. For \(a=0.5\), \(p=50\,M\), \(e=0.1\), and \(z_1=1/\sqrt{2}\), we obtain
\begin{equation}
    T_{\mathrm{LT}}^{\mathrm{th}}
    =\frac{2\pi}{|\Omega_\phi-\Omega_\theta|}
    =8.278\times10^5\,M.
\end{equation}
The inferred nodal-precession period is in good agreement with the theoretical value. This shows that the Lense-Thirring precession period is encoded not only in QPE timing~\citep{Zhou:2025auz}, but also in the self-lensing envelope. The same geodesic also gives the theoretical orbital period
\begin{equation}
    T_{\mathrm{orb}}^{\mathrm{th}}
    =\frac{2\pi}{\Omega_\theta}
    =2.261\times10^3\,M.
\end{equation}

A closer view of one envelope is shown in Panel~(b), where the even branch is followed across the strong-lensing sector from $\phi_{o,\mathrm{rel}}^{\mathrm{even}}\simeq195^\circ$ to $165^\circ$. Across this interval, the flare amplitudes vary substantially as the source passes through the high-magnification region. By contrast, the recurrence time within each branch remains nearly constant,
\begin{equation}
    \langle T_{\mathrm{odd}}\rangle
    \simeq
    \langle T_{\mathrm{even}}\rangle
    \simeq
    \langle T_{\mathrm{long}}+T_{\mathrm{short}}\rangle
    \simeq
    T_{\mathrm{orb}}^{\mathrm{th}} \ .
\end{equation}
Thus, the local timing remains consistent with the EMRI apsidal precession interpretation of an alternating long-short recurrence pattern~\citep{CongZhou2024a,CongZhou2024b,secularEvolutionQPE2025,Zhou:2025auz,Zhou:2026iub,Zeng:2026ajf}, while self-lensing primarily modulates the observed amplitudes.

Panel~(c) illustrates the light curve over a shorter observing window containing only a few flares. On this timescale, the light curve displays both the long-short recurrence pattern and a strong-weak amplitude pattern. The former reflects the intrinsic timing of consecutive disk crossings, whereas the latter arises naturally from the changing lensing geometry. Neighboring events sample different source-lens-observer alignments: one event may occur closer to the far-side strong-lensing configuration, while the next is dominated by the weakly lensed direct image. Self-lensing therefore provides a relativistic mechanism for producing a local strong-weak flare pattern without requiring event-to-event changes in the intrinsic flare emission ~\citep{multibandstardisksimulation2025, Jankovi2026A&A...710A.405J, Jankovi2026arXiv260705508J, huang2026ApJ..1005..235H, wrap_disk_QPE_BRD_2026}. This interpretation is clarified by the image-component decomposition in Panel~(d). Near-side events are dominated by the direct image, with magnification of order unity. For far-side events, the observed flare is the sum of magnified, time-delayed copies of the same intrinsic flare. Their different magnifications and propagation delays determine both the total flare amplitude and the detailed temporal profile. In the cases considered here, the observed signal is well described by the low-order \(N=0,1\) images. Higher-order images are more delayed but much more strongly demagnified, so their contributions do not noticeably affect the light curve~\citep{Forward_RTracing_2025,Lensing_by_KerrBH_2020}.  For simplicity, we have assumed that both images of  \(N=0,1\) reach the observer. In fact, the accretion disk may block one of the images in some geometry, which makes the lensing light curve prediction less certain. But the behavior of two branches and long-term modulation in the light curve is unaffected.

\subsection{Magnification Maps and Lensing Probabilities}
\label{subsec:lensing_possibility}

The self-lensed light curves discussed above show that the observed flare amplitudes depend sensitively on the instantaneous observer-SMBH-source geometry. To quantify this dependence, we construct point-source magnification maps on the observer sky. For a source fixed on the equatorial plane at radius \(r_s\), each viewing direction \((\theta_o,\phi_{o,\mathrm{rel}})\) defines a distinct lensing configuration. We ray trace the retained image components and calculate the total point-source magnification as
\begin{equation}
    \mu_{\mathrm{tot}}(\theta_o,\phi_{o,\mathrm{rel}};r_s)
    =
    \sum_{\ell}
    \mu_{\ell}(\theta_o,\phi_{o,\mathrm{rel}};r_s),
    \label{eq:total_magnification_map}
\end{equation}
where \(\ell\) labels the image component. Since the delays between the dominant low-order images are short compared with the intrinsic flare duration in the configurations considered here, their summed magnification approximates the enhancement of the flare amplitude.

Figure~\ref{fig:lensing_possibility_maps} shows that the magnification is concentrated around the nearly aligned far-side configuration, \(\theta_o\simeq90^\circ\) and \(\phi_{o,\mathrm{rel}}\simeq180^\circ\). The rapid decline away from alignment restricts appreciable self-lensing to a small region of the observer sky. The three-dimensional surface for \(r_s=50\,M\) highlights the steep central spike near a point-source caustic. Its displayed height is a numerical truncation of the formal divergence; averaging over a finite emitting region, which is not included here, would regularize the peak.

To quantify the angular coverage of self-lensing, we define the region above a magnification threshold \(\mu_{\mathrm{th}}\) as
\begin{equation}
    \mathcal{R}_{\mu_{\mathrm{th}}}(r_s)
    =
    \left\{
    (\theta_o,\phi_{o,\mathrm{rel}})
    \,\big|\,
    \mu_{\mathrm{tot}}(\theta_o,\phi_{o,\mathrm{rel}};r_s)>
    \mu_{\mathrm{th}}
    \right\},
    \label{eq:lensing_region}
\end{equation}
with the corresponding observer-sky solid angle
\begin{equation}
    \Delta\Omega_{\mu_{\mathrm{th}}}(r_s)
    =
    \int_{\mathcal{R}_{\mu_{\mathrm{th}}}(r_s)}
    \sin\theta_o\,\mathrm{d}\theta_o\,
    \mathrm{d}\phi_{o,\mathrm{rel}}.
    \label{eq:lensing_solid_angle}
\end{equation}
For isotropically distributed viewing directions, the geometric lensing probability for a single flare at this source position is
\begin{equation}
    P_{\mu_{\mathrm{th}}}(r_s)
    =
    \frac{\Delta\Omega_{\mu_{\mathrm{th}}}(r_s)}{4\pi}.
    \label{eq:lensing_probability}
\end{equation}

The white contours in Figure~\ref{fig:lensing_possibility_maps} mark two representative thresholds. We adopt \(\mu_{\mathrm{th}}=1.2\) for the modest enhancement that produces a visible strong-weak pattern in Figure~\ref{fig:Static_LightCurve}, while \(\mu_{\mathrm{th}}=5\) identifies the strongly amplified core. Integrating the region above the fiducial threshold at \(r_s=50\,M\) gives
\begin{equation}
    \begin{aligned}
        \Delta\Omega_{1.2}(50\,M)
        &\simeq 0.39\,\mathrm{sr},\\
        P_{1.2}(50\,M)
        &\simeq 3.1\%.
    \end{aligned}
    \label{eq:lensing_probability_fiducial}
\end{equation}

The probability above refers to a single flare at a fixed source position. Over a longer observation, Lense-Thirring precession shifts the crossing sites in azimuth, causing their above-threshold regions to sweep across the observer sky. The cumulative lensing probability \(P_{\mu_{\mathrm{th}}}^{(\mathrm{c})}\) is the fraction of viewing directions from which at least one flare exceeds the threshold during the observing window.

For two approximately antipodal crossing branches at the same radius, the regions enclosed by their \(\mu_{\mathrm{th}}\) contours together sweep out the full equatorial belt within half a precession cycle, giving
\begin{equation}
    P_{\mu_{\mathrm{th}}}^{(\mathrm{c})}(T_{\mathrm{obs}};r_s)
    =P_{\mathrm{belt}}(r_s)
    \quad ({\rm for}\ T_{\mathrm{obs}}\geq T_{\mathrm{LT}}/2) .
    \label{eq:cumulative_lensing_probability}
\end{equation}
Here, \(T_{\mathrm{LT}}\) is the full nodal-precession period, and \(P_{\mathrm{belt}}\) is the fraction of the sky covered by this belt at the same magnification threshold. Figure~\ref{fig:lt_lensing_probability} shows how the cumulative lensing probability increases with observing duration for the four source radii. For \(r_s=50\,M\) and \(\mu_{\mathrm{th}}=1.2\), it reaches \(P_{\mathrm{belt}}=34.8\%\) by half a precession period.

The radius dependence of the single-flare probability is summarized in Figure~\ref{fig:lensing_possibility_scaling}. The nearly horizontal curves show that the solid-angle ratios depend only weakly on the magnification threshold: increasing \(r_s\) reduces both the modestly and strongly amplified regions by similar factors. Averaged over the threshold range shown, the relative solid angles at \(r_s=100\), \(150\), and \(200\,M\), normalized to the \(50\,M\) value, are approximately \(0.491\), \(0.322\), and \(0.237\), respectively, close to the inverse-radius sequence \(1/2\), \(1/3\), and \(1/4\). Over the sampled radii and thresholds, these ratios are approximately described by
\begin{equation}
    \frac{\Delta\Omega_{\mu_{\mathrm{th}}}(r_s)}
         {\Delta\Omega_{\mu_{\mathrm{th}}}(50\,M)}
    \simeq
    \left(\frac{r_s}{50\,M}\right)^{-1}.
    \label{eq:lensing_solid_angle_scaling}
\end{equation}

This approximate inverse-radius behaviour follows from the small-angle point-lens limit. Let \(\delta\) denote the angular separation from the far-side alignment, \((\theta_o,\phi_{o,\mathrm{rel}})=(90^\circ,180^\circ)\). For an observer at \(R\equiv r_o\gg r_s\), the unlensed source angle is \(\beta\simeq r_s\delta/R\), while the Einstein angle is \(\theta_E\simeq(4Mr_s)^{1/2}/R\). The dimensionless impact parameter is therefore
\begin{equation}
    y\equiv\frac{\beta}{\theta_E}
    \simeq
    \frac{\delta}{2}
    \left(\frac{r_s}{M}\right)^{1/2}.
\end{equation}
A fixed magnification threshold corresponds to a fixed limiting value of \(y\), so the angular radius of the threshold contour scales as \(\delta_{\mathrm{th}}\propto r_s^{-1/2}\), giving
\begin{equation}
    \Delta\Omega_{\mu_{\mathrm{th}}}
    \sim\pi\delta_{\mathrm{th}}^2
    \propto r_s^{-1}.
    \label{eq:lensing_solid_angle_scaling_argument}
\end{equation}

Since \(P_{\mu_{\mathrm{th}}}=\Delta\Omega_{\mu_{\mathrm{th}}}/(4\pi)\), the single-flare probability follows the same scaling. At the fiducial threshold, this gives
\begin{equation}
    P_{1.2}(r_s)
    \simeq
    3.1\%
    \left(\frac{r_s}{50\,M}\right)^{-1}.
    \label{eq:lensing_probability_scaling}
\end{equation}
Thus, individual flares originating closer to the SMBH have a larger geometric probability of exceeding the adopted threshold. For \(T_{\mathrm{obs}}\geq T_{\mathrm{LT}}/2\), the cumulative probability is set by the angular width of the equatorial belt. The same \(\delta_{\mathrm{th}}\propto r_s^{-1/2}\) scaling therefore gives, in the small-angle limit,
\begin{equation}
    P_{1.2}^{(\mathrm{c})}(T_{\mathrm{obs}};r_s)
    \simeq
    34.8\%
    \left(\frac{r_s}{50\,M}\right)^{-1/2}.
    \label{eq:cumulative_lensing_probability_scaling}
\end{equation}

\begin{figure*}[htb]
    \centering
    \includegraphics[width=\textwidth]{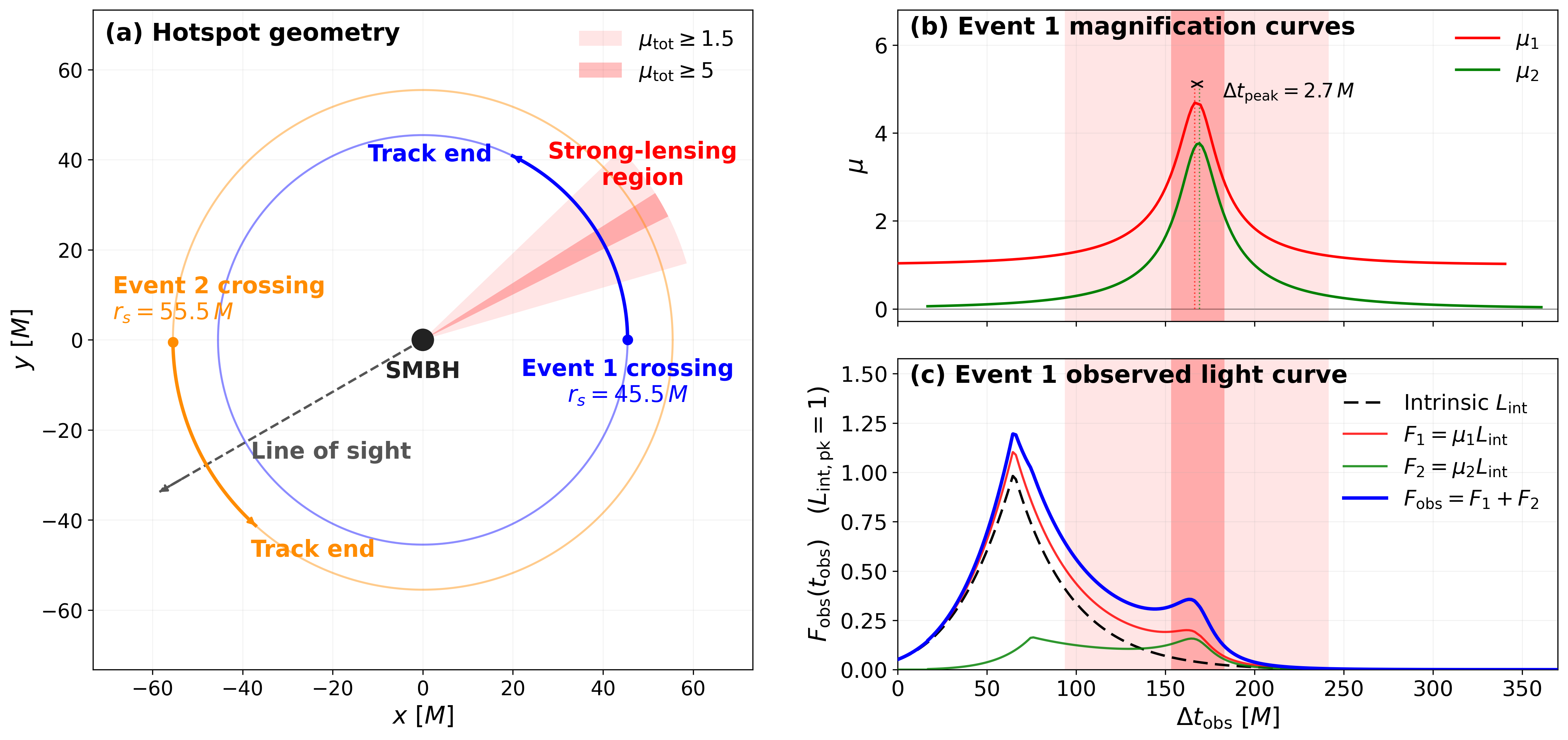}
    \caption{ Lensing signatures of Keplerian hotspots following two adjacent disk-crossing events of the fiducial EMRI orbit, viewed at \(\theta_o=88^\circ\). Panel~(a) shows the equatorial-plane geometry. The light- and dark-red regions mark \(\mu_{\mathrm{tot}}\geq1.5\) and \(\mu_{\mathrm{tot}}\geq5\), respectively. Since the EMRI orbit is eccentric, Events~1 and~2 occur at different radii, \(r_s=45.5\,M\) and \(55.5\,M\), respectively. The Event~1 hotspot starts on the far side and moves through the strongly magnified region. By contrast, the Event~2 hotspot starts on the near side and remains direct-image dominated. Panel~(b) shows the time-dependent magnifications of the \(N=0\) and \(N=1\) lensed images for Event~1. The inter-image delay also varies as the hotspot moves, being larger near the beginning and end of the curves and smaller in between. The two magnification peaks are separated by approximately \(2.7\,M\) in observer time. Panel~(c) compares the intrinsic flare, the two image contributions, and  total observed light curve. The peak of the evolving magnification occurs after the intrinsic luminosity peak, enhancing the declining wing and producing a trailing shoulder.}
    \label{fig:rotation-lightcurve-comparison}
\end{figure*}

\section{Self-Lensing of Keplerian Hotspots}
\label{sec:rotating}

The source prescriptions introduced in Section~\ref{subsec:forward_raytracing} differ in whether the emitting region remains fixed or moves during a flare. Here, we study the Keplerian prescription. Following a disk crossing, the shocked gas expands and cools, and radiation escapes as its optical depth decreases~\citep{sBH_disk2026}. We represent the luminous region by a pointlike hotspot. We assume that the emitting region co-rotates azimuthally at the local prograde Keplerian rate during the radiative phase. Therefore, the source-lens-observer geometry evolves within a single flare, resulting in time-dependent image magnifications and photon propagation times. Hotspot motion therefore introduces an additional, time-dependent lensing response beyond the fixed superposition of delayed images in the static model.

Throughout this section, we retain the fiducial black-hole and orbital parameters adopted in Section~\ref{sec:static}: \(a=0.5\), \(p=50\,M\), \(e=0.1\) and \(z_1=1/\sqrt{2}\). We adopt a more nearly edge-on viewing inclination of \(\theta_o=88^\circ\) to illustrate the intra-flare distortions produced by hotspot motion. The dependence of lensing strength on source radius and observer inclination was quantified in Section~\ref{sec:static}. We use the same intrinsic flare prescription as in the static calculation. As defined in Section~\ref{subsec:intrinsic_flare}, \(\tilde t=0\) denotes the disk crossing at \(t_e=t_i\). We adopt \(t_p=60\,M\) and \(\tau_2=30\,M\).

\subsection{Self-Lensed Light Curve of a Keplerian Hotspot}
\label{subsec:keplerian_lightcurve}

We use the Keplerian source trajectory defined by Equations~\eqref{eq:source_trajectory_local_time} and \eqref{eq:source_angular_velocity}. We model the hotspot moving in a circular orbit with a local prograde Keplerian rate. Since the EMRI orbit is eccentric, the two representative events in Figure~\ref{fig:rotation-lightcurve-comparison} occur at different radii, \(r_s=45.5\,M\) and \(55.5\,M\), respectively.

Figure~\ref{fig:rotation-lightcurve-comparison}(a) compares two adjacent disk crossings from the interleaved branches defined in Section~\ref{subsec:qpe_event_sequence}. Event~1 on the far side moves through the strongly magnified region. Along the Event~1 trajectory, the light-red region marks \(\mu_{\mathrm{tot}}=\mu_1+\mu_2\geq1.5\), while the nested dark-red region marks \(\mu_{\mathrm{tot}}\geq5\). Event~2  on the near side does not experience strong lensing.

The observed signal is calculated with the emission-to-arrival-time mapping in Equations~\eqref{eq:image_observer_time_mapping} and \eqref{eq:image_flux_parametric}. We retain the two dominant image branches and refer to the \(N=0\) and \(N=1\) branches as the primary and secondary images, respectively. Taking Event 1 as an example, as the hotspot moves during a single flare, the magnifications of both lensed images vary with time, as shown in Figure~\ref{fig:rotation-lightcurve-comparison}(b). Notably, the inter-image delay also varies as the lensing geometry changes along the hotspot trajectory, being larger near the beginning and end of the curves and smaller in between. In this example, the primary magnification peak precedes the secondary peak by approximately \(2.7\,M\) in observer time.

For Event~1, the intrinsic flare reaches its maximum before the moving hotspot enters the most strongly magnified portion of its trajectory. The late magnification episode therefore amplifies the declining wing of the intrinsic flare, producing the trailing shoulder visible in Figure~\ref{fig:rotation-lightcurve-comparison}(c). This shoulder is not a delayed copy of the intrinsic flare peak in the secondary image. Instead, it arises as the time-dependent magnifications reach their maxima during the flare's decline. By contrast, Event~2 on the near side remains direct-image dominated, and its observed profile stays close to the intrinsic flare. As Lense--Thirring precession changes the crossing azimuth, the magnification episode shifts relative to the intrinsic peak in successive flares. An extended example is shown in Appendix~\ref{subsec:keplerian_evolution}; the consequences for the effective flare width are discussed below.

\subsection{Distinguishing Keplerian Hotspots from Static Hotspots}
\label{subsec:static_keplerian_comparison}

For a static hotspot, each lensed image is a delayed copy of the intrinsic flare with a constant magnification, as shown in Figure~\ref{fig:Static_LightCurve}(d). By contrast, a Keplerian hotspot moves while it emits, so both the magnification and photon propagation time change during the flare. Different stages of the intrinsic emission can therefore be magnified by different amounts, changing the observed profile.

The trailing shoulder in Figure~\ref{fig:rotation-lightcurve-comparison}(c) illustrates this effect: the strongest magnification occurs after the intrinsic luminosity peak and enhances the declining emission. Whether this enhancement produces a shoulder or a separate second peak depends on its strength and timing. The example shown has \(r_s\simeq50\,M\) and a nearly edge-on viewing angle. However, for hotspots farther from the SMBH, or observers farther from the equatorial plane, the maximum magnification can be lower, as shown in Appendix~\ref{appsubsec:magnification_envelopes} (Figure~\ref{fig:envelope_profile_comparison}). In these configurations, hotspot motion may still change the flare width or asymmetry without producing a visible second peak. Searching only for double-peaked flares would therefore miss these less pronounced effects.

\begin{figure}[htb]
    \centering
    \includegraphics[width=\columnwidth]{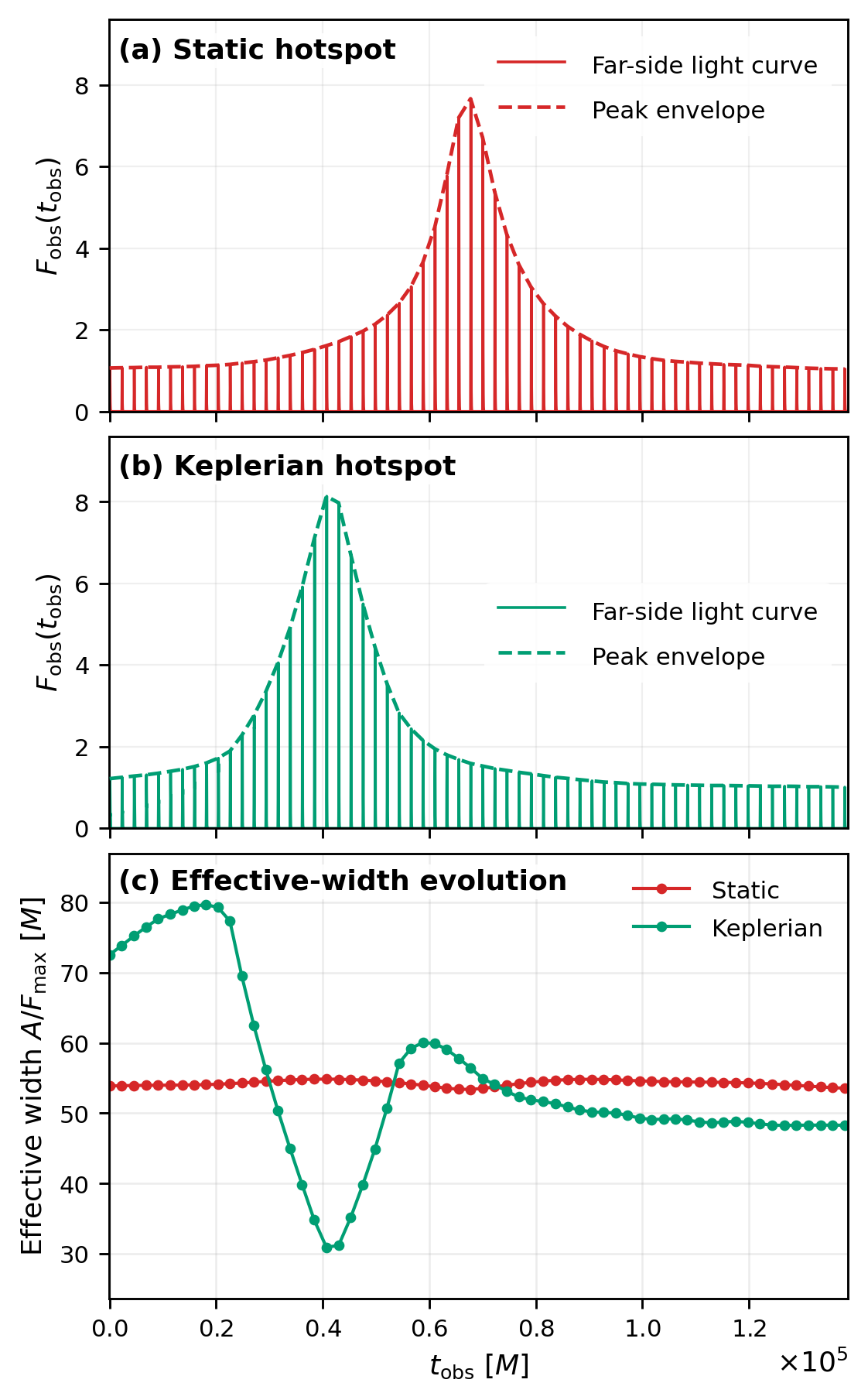}
    \caption{Comparison of far-side flares in the static and Keplerian hotspot models. Both calculations use \(\theta_o=88^\circ\), the same disk-crossing sequence, and the same intrinsic flare parameters. Panels~(a) and (b) show the far-side light curves and their peak envelopes for the static and Keplerian cases, respectively. The time offset between the envelope maxima depends on the chosen initial phase and does not by itself distinguish the two models. Panel~(c) compares the effective widths, \(W_{\mathrm{eff}}=A/F_{\mathrm{max}}\). The width remains nearly constant in the static example, whereas it varies in the Keplerian example as the strongest magnification shifts relative to the intrinsic luminosity peak.}
    \label{fig:effective-width}
\end{figure}

To quantify changes in flare shape even when no separate second peak is visible, we define the effective width of an individual flare:
\begin{equation}
    W_{\mathrm{eff}}
    \equiv
    \frac{A}{F_{\mathrm{max}}},
    \qquad
    A
    \equiv
    \int_{\text{flare}} F_{\mathrm{obs}}(t_{\mathrm{obs}})\,\mathrm{d}t_{\mathrm{obs}},
\label{eq:effective_flare_width}
\end{equation}
where \(A\) is the observed fluence of a single disk-crossing event and \(F_{\mathrm{max}}\) is the peak flux. Equivalently, \(W_{\mathrm{eff}}\) is the width of a rectangular flare with the same fluence and peak height.

Panels~(a) and (b) of Figure~\ref{fig:effective-width} compare the far-side flare sequences in the two models. In both cases, precession changes the flare amplitudes as the crossing sites move through the strongly magnified region. Their effective widths, however, behave differently. As shown in Panel~(c), \(W_{\mathrm{eff}}\) remains close to \(54\,M\) in the static example, whereas it varies from approximately \(30\) to \(80\,M\) in the Keplerian example. The Keplerian width evolution is also asymmetric about the phase of strongest lensing.

The width increases or decreases according to which part of the flare receives the strongest magnification. When lensing mainly enhances emission before or after the main peak, the fluence can increase by a larger fraction than the peak flux, increasing \(W_{\mathrm{eff}}\). If the strongest magnification instead occurs near the intrinsic peak, the peak flux can increase by a larger fraction than the fluence, reducing \(W_{\mathrm{eff}}\). As precession changes the initial hotspot position, successive flares can therefore become either broader or narrower in this measure. Figure~\ref{fig:evolution_keplerian_Lightcurve} in Appendix~\ref{subsec:keplerian_evolution} illustrates how the magnification peak shifts relative to the intrinsic luminosity peak, reshaping individual flares and changing their effective widths. Moreover, because the intrinsic flare rises faster than it decays, magnification at equal time offsets before and after the peak need not produce equal width changes. This helps explain the asymmetry of the width evolution.

In our model, the widths and shapes of far-side flares evolve while the alternating near-side profiles remain comparatively stable. Tracking this pattern can therefore help identify hotspot motion even when the light curves do not show a separate second peak.

\subsection{Distinguishing Self-Lensing from External Lensing}
\label{subsec:self_external_lensing}

The double-peaked light curve is not unique to self-lensing of Keplerian hotspots: external gravitational lensing can produce similar profiles for suitable parameter choices \citep{overlapping_LightCurve_by_External_lensing2026}. In our model, the later peak arises from varying lensing magnification as the hotspot's azimuth evolves (Fig.~\ref{fig:evolution_keplerian_Lightcurve}), whereas an external lens can produce a second peak through the delayed arrival of another lensed image of the same flare.

The key distinction is when the photons contributing to the later peak were emitted. In our self-lensing example, this peak samples a later, cooler stage of the flare. A delayed image produced by an external lens instead repeats the intrinsic spectral evolution of the original flare. The observed peak spectra can nevertheless differ if the two images overlap in time, as illustrated below.

\begin{figure*}[htb]
    \centering
    \includegraphics[width=\textwidth]{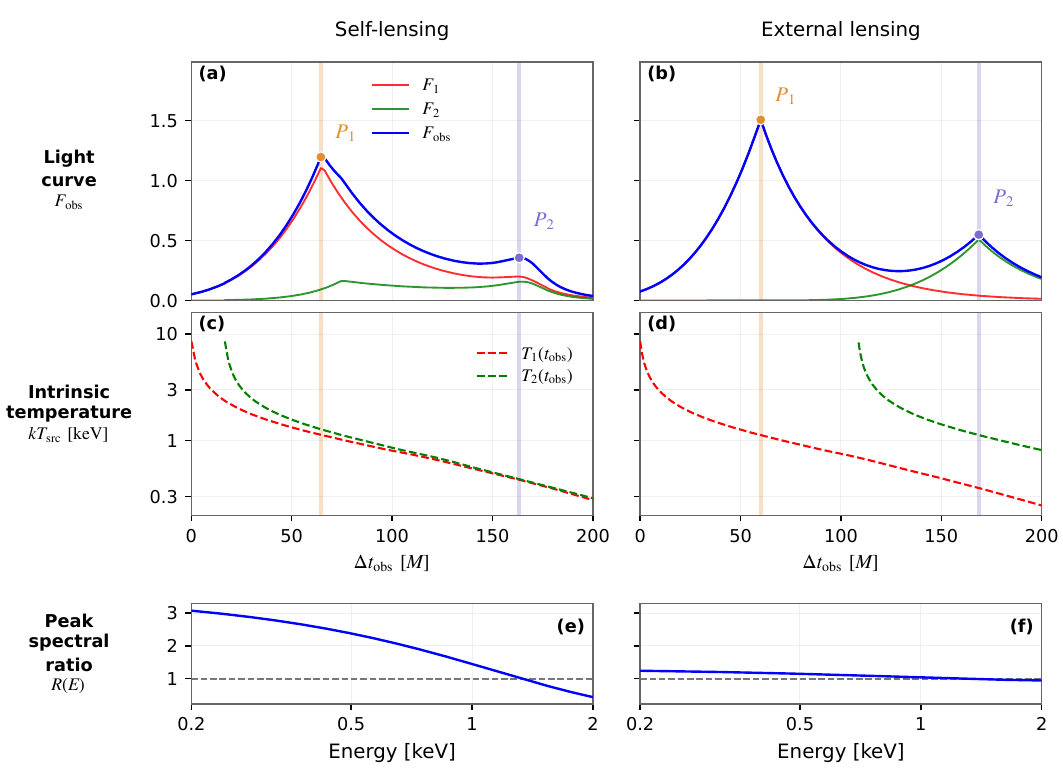}
    \caption{The comparison of self-lensing (left) and external lensing (right) with similar double-peaked light curves. Panels~(a) and (b) show the total observed flux, $F_{\rm obs}=F_1+F_2$ (blue), with the two constituent images $F_1$ and $F_2$ shown in red and green, respectively. Both cases use a common flux scale with $L_p=1$. Orange and purple bands in the upper two rows mark the spectral extraction windows centred on the total-flux peaks $P_1$ and $P_2$, respectively. Panels~(c) and (d) show the intrinsic temperature evolution of the plasma ball (hotspot) produced by a disk-crossing event. The curves $T_1(t_{\rm obs})$ (red dashed) and $T_2(t_{\rm obs})$ (green dashed) give its temperature at the emission times associated with the two lensed images, plotted against observer time. The later self-lensing peak samples a later, cooler stage of the plasma ball's evolution, whereas the delayed external-lensing image reproduces the same cooling history with a time offset. Panels~(e) and (f) show $R(E)=\widehat{F}_{E,P_2}/\widehat{F}_{E,P_1}$, the ratio of the time-averaged total spectra around the two peaks after each is normalised to unit integrated flux over $0.2$-$2.0\,\mathrm{keV}$. The dashed $R=1$ line corresponds to identical spectral shapes. The self-lensing example exhibits stronger spectral softening. The smaller departure from unity in the external-lensing example reflects the first image's cooling-tail contribution at $P_2$. }
    \label{fig:self-external-spectral}
\end{figure*}

Figure~\ref{fig:self-external-spectral} compares the two lensing scenarios using the same prescribed intrinsic flare from a plasma ball produced by a disk-crossing event. The band luminosity follows Equation~\eqref{eq:phenom_intrinsic_flare}, with $t_p=60\,M$, $\tau_2=30\,M$, and $L_p=1$. At each emission time, we scale the blackbody spectrum at the plasma ball's intrinsic temperature to match the prescribed $0.2$-$2.0\,\mathrm{keV}$ band luminosity. The temperature and luminosity histories are specified independently. The cooling history is adapted from Figure~9 of \citet{CongZhou2024a}, with its $0$-$4000\,\mathrm{s}$ interval rescaled to $0$-$200\,M$ for illustration. Under our adopted approximation, lensing changes magnifications and arrival times without shifting photon energies, so differences between the peak spectra reflect the different stages of the plasma ball's intrinsic cooling sampled by the lensed images.

For the external-lensing example, we adopt a static point-mass lens with dimensionless source offset $u\equiv\beta/\theta_{\rm E}=0.55$ and redshifted mass $M_{{\rm L},z}\equiv(1+z_{\rm L})M_{\rm L}=24.4\,M$. Here, $\beta$ is the angular source-lens separation, $\theta_{\rm E}$ is the Einstein angle, $M_{\rm L}$ and $z_{\rm L}$ are the lens mass and redshift, and $M$ is the central SMBH mass. These parameters give constant image magnifications $\mu_+\simeq1.509$ and $|\mu_-|\simeq0.509$, with an inter-image delay $\Delta t_{\rm img}\simeq108.7\,M$.

To compare the peak spectra, we average the total spectrum, including both image contributions, over a $2\,M$-wide observer-time window centred on each peak $P_k$ ($k=1,2$). Denoting this average by $\langle F_E^{\rm obs}\rangle_{P_k}$, we define the normalised spectra and their ratio as
\begin{align}
    \widehat{F}_{E,P_k}
    &=
    \frac{\langle F_E^{\rm obs}\rangle_{P_k}}
    {\displaystyle\int_{0.2\,{\rm keV}}^{2.0\,{\rm keV}}
    \langle F_{E'}^{\rm obs}\rangle_{P_k}\,\mathrm{d}E'},
    \label{eq:peak_spectrum_normalization}
    \\
    R(E)
    &\equiv
    \frac{\widehat{F}_{E,P_2}}{\widehat{F}_{E,P_1}}.
    \label{eq:peak_spectral_ratio}
\end{align}
This normalisation removes the overall brightness difference, so $R(E)=1$ throughout the band corresponds to identical spectral shapes.

The self-lensing example shows a much stronger decrease of $R(E)$ with energy, indicating greater softening of the second peak. The smaller departure from unity in the external-lensing example reflects the contribution of the first image's cooling tail at the delayed peak. This comparison uses the total spectra of the unresolved images, and its diagnostic power depends on the intrinsic spectral evolution and the degree of image overlap.

\section{Discussion and Conclusions}
\label{sec:discuss}
\label{sec:con}

In this work, we have studied the self-lensing of electromagnetic flares from EMRI-disk crossings using forward ray tracing in Kerr spacetime. We find that self-lensing can produce strong-weak alternation even when successive disk crossings generate identical intrinsic emission. For suitable viewing directions, this pattern arises because far-side flares are strongly magnified whereas near-side flares remain almost unchanged. As Lense-Thirring precession shifts the crossing positions, the flare amplitudes vary over time and form broad envelopes. In our model, the two approximately opposite crossing branches reach their strongest magnification in turn. Neighbouring envelope maxima in the combined sequence are therefore separated by about half a nodal-precession period, allowing that period to be inferred from their spacing.

The same precession can also increase the cumulative geometric lensing probability as the observing duration increases. As the crossing sites move around the SMBH, flares can become strongly magnified for viewing directions that initially showed little magnification. We quantify this effect by the fraction of directions from which at least one flare exceeds a chosen magnification threshold. This fraction is larger for flares produced closer to the SMBH, where appreciable magnification extends over a wider range of directions. In practice, however, detecting these flares also depends on their brightness, observing cadence, and absorption by surrounding gas, especially near the disk plane.

While Lense-Thirring precession changes the lensing geometry between successive flares, Keplerian motion changes it during an individual flare. Our results show that the flare profile depends on when the strongest magnification occurs relative to the intrinsic luminosity peak. When magnification is strongest after the intrinsic peak, it enhances the fading emission and can produce a shoulder or second peak. Conversely, magnification near the intrinsic peak preferentially increases the peak flux. The effective width can thus either increase or decrease, depending on which part of the flare is amplified. Because precession changes the initial hotspot position, these distortions evolve from one flare to the next. In our examples, far-side flares show this evolution while alternating near-side flares remain close to the intrinsic profile. Changes in width and asymmetry can therefore help identify hotspot motion even without a resolved second peak.

Within the EMRI-disk scenario, self-lensing provides an additional mechanism for producing the strong-weak pattern seen in some QPEs.  The cumulative geometric probabilities found here indicate that modest amplification can occur for a substantial range of viewing directions over longer observations. This suggests that self-lensing may already contribute to some observed QPE light curves and should be considered when interpreting their amplitude patterns. An external-lensing interpretation has been proposed for the overlapping eruptions in eRO-QPE1~\citep{overlapping_LightCurve_by_External_lensing2026}. However, our calculations show that self-lensing of a moving hotspot can also produce double-peaked profiles, so light-curve morphology alone does not establish an external-lensing origin. Self-lensing therefore provides a physically motivated alternative that merits testing against the same observations.

To distinguish self-lensing from external lensing, we examine the spectral evolution associated with the two peaks. In our illustrative cooling model, we find that self-lensing produces a softer second peak by amplifying later, cooler emission. By contrast, each image formed by a static external lens repeats the same intrinsic spectral evolution. This distinction becomes less clear when the images overlap, because the first image's cooling tail also contributes to the second peak. Our comparison therefore suggests a possible spectral diagnostic, whose usefulness depends on the cooling history and the degree of image overlap. Since our luminosity and temperature histories are prescribed independently, quantitative predictions require a cooling model that determines both.

More generally, our calculation focuses on geometric lensing and represents the emitting gas by point sources with prescribed intrinsic flare profiles. For sufficiently compact emitters, finite-source averaging is expected to preserve strong-weak alternation, precession-driven amplitude envelopes, and the late-time enhancement that Keplerian motion can produce. Peak amplitudes and the prominence of shoulders or resolved secondary peaks may nevertheless change, especially near caustics. We include geometric magnification in the flux calculation but neglect Doppler and gravitational frequency shifts, their effects on intensity, and absorption. We also neglect long-term orbital evolution driven by GW emission or disk interactions. Quantitative fits would need to account for the size and evolution of the emitting gas and assess the importance of the omitted effects.


\begin{acknowledgments}
R.Z. thanks Yuhe Zeng, Binyu Zhong, and Zirui Zhang for useful discussions.
Z.P. thanks the support by the National Natural Science Foundation of China (grants  12673052).
The computations in this paper were run on the cluster supported by the Astronomy and Astrophysics Division of Tsung-Dao Lee Institute, Shanghai Jiao Tong University.
\end{acknowledgments}

\begin{figure*}[htb]
\centering
\includegraphics[width=\textwidth]{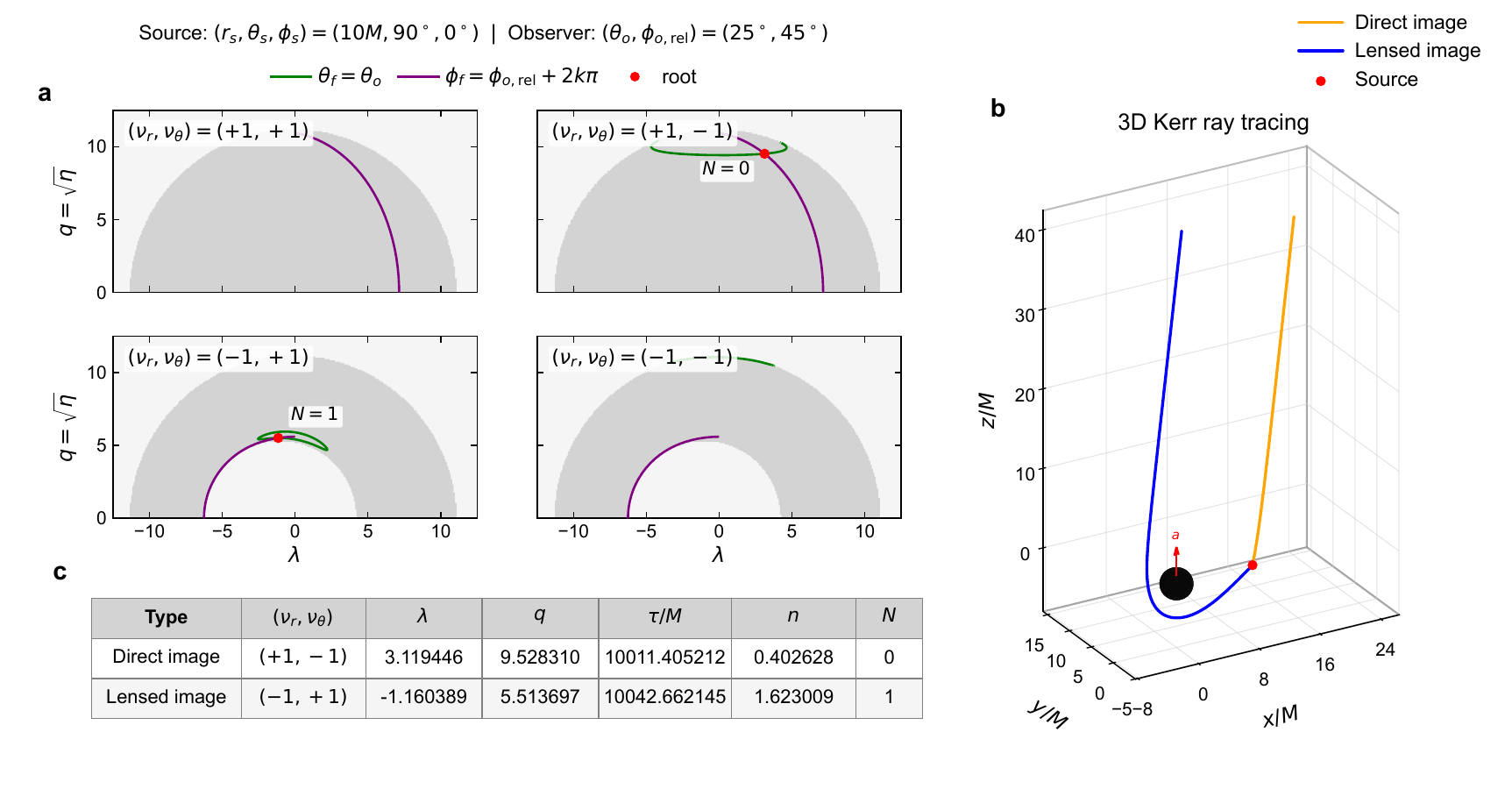}
\caption{ Low-order images found directly in the \((\lambda,q)\) plane for an equatorial source at \(r_s=10\,M\) around a Kerr black hole with dimensionless spin \(a=0.5\), observed from \((\theta_o,\phi_{o,\mathrm{rel}})=(25^\circ,45^\circ)\) and \(r_o=10^4\,M\). \textbf{(a)} Green and purple contours show \(\theta_f=\theta_o\) and \(\phi_f=\phi_{o,\mathrm{rel}}+2\pi k\), respectively, for the four initial-sign branches. Their intersections (red points) identify the \(N=0\) and \(N=1\) images. \textbf{(b)} The corresponding null geodesics near the black hole. The direct and lensed rays are shown in orange and blue, with the source in red and the horizon in black. \textbf{(c)} The initial signs, conserved quantities, propagation times, polar half-orbit counts, and image orders of the two solutions. }
\label{fig:lambda_q_rootfinding}
\end{figure*}
\begin{figure}[htb]
    \centering
    \includegraphics[width=\columnwidth]{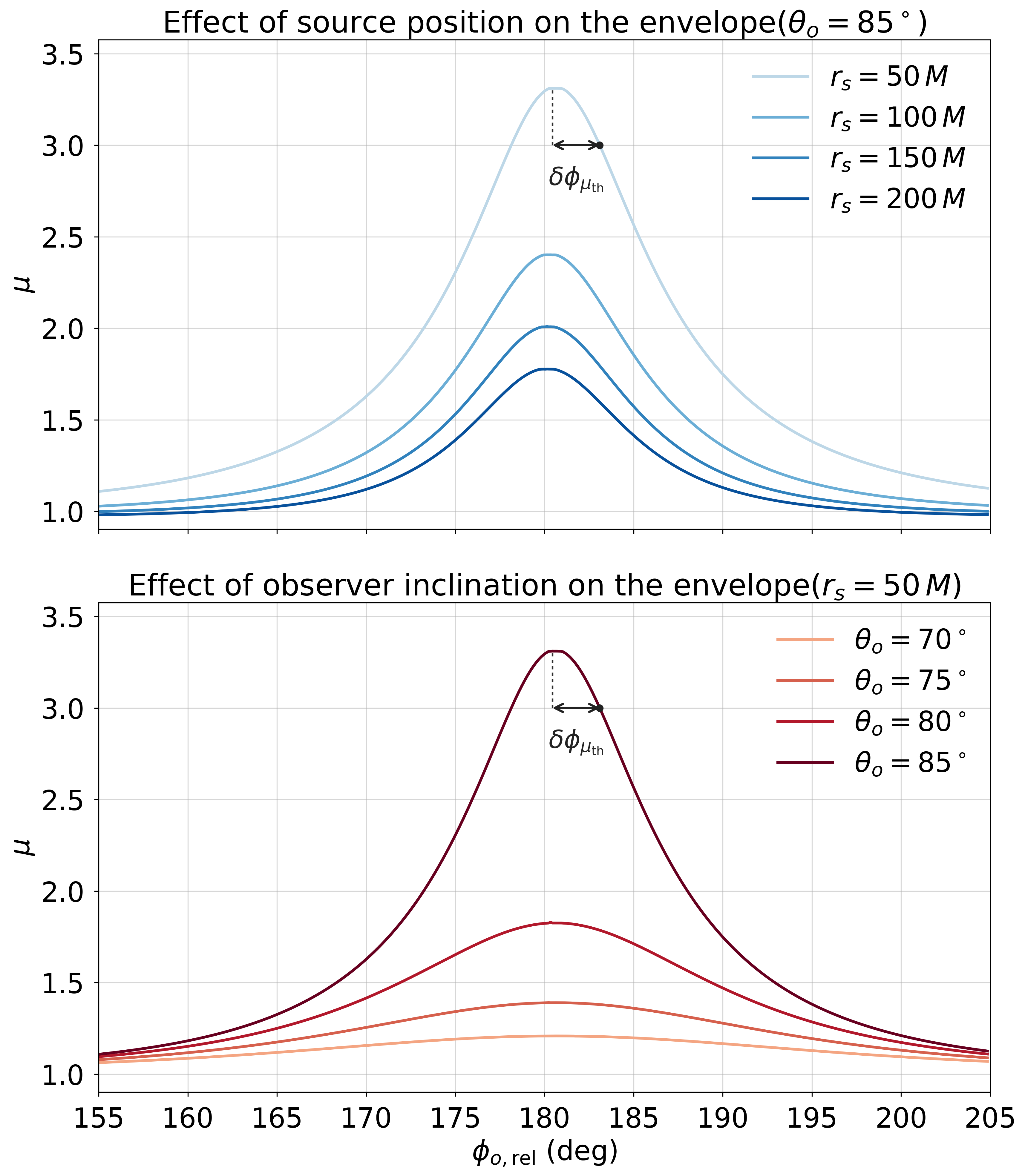}
    \caption{ One-dimensional magnification profiles extracted from the maps in Figure~\ref{fig:lensing_possibility_maps}. \textbf{Top:} fixed-inclination cuts at \(\theta_o=85^\circ\) for \(r_s=50,100,150,\) and \(200\,M\). The peak magnification decreases with increasing source radius. \textbf{Bottom:} fixed-radius cuts at \(r_s=50\,M\) for \(\theta_o=70^\circ,75^\circ,80^\circ,\) and \(85^\circ\). Moving the observer closer to the equatorial plane raises the peak and widens the interval above a fixed magnification threshold. Lense-Thirring precession causes successive disk crossings to sample these profiles, producing the slow modulation of flare amplitudes. The black arrows mark the measured one-sided half-widths for \(\mu_{\mathrm{th}}=3.0\), from the actual profile maximum to the right-hand threshold crossing; Equation~\eqref{eq:lensing_cut_half_width} gives their circular-contour approximation. }
    \label{fig:envelope_profile_comparison}
\end{figure}
\begin{figure*}[htb]
    \centering
    \includegraphics[width=\textwidth]{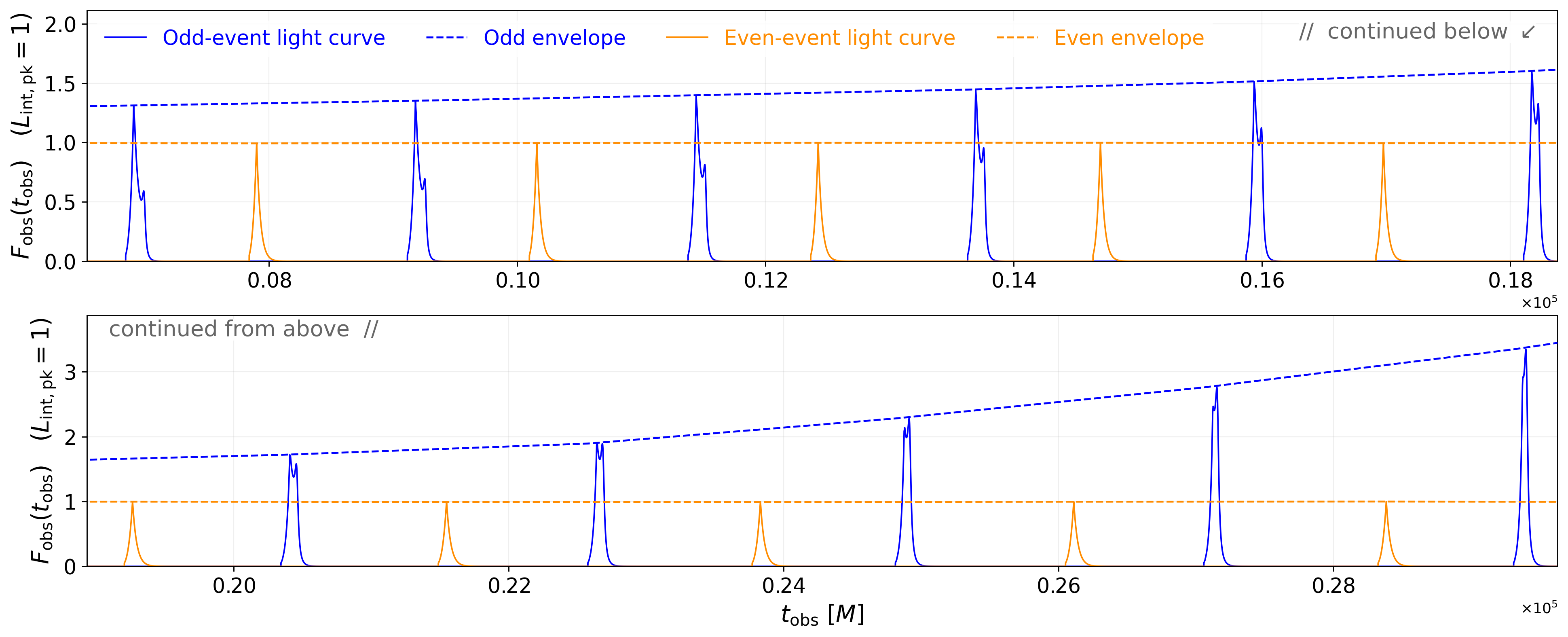}
    \caption{ Evolution of individual flares in the Keplerian-hotspot model. A single flare sequence is split across the upper and lower panels. In the sequence shown, odd events are far-side flares and even events are near-side flares. Along each crossing branch, Lense--Thirring precession changes the initial hotspot phase from one event to the next. For the far-side flares, the magnification peak moves progressively closer in time to the intrinsic luminosity peak, shifting the late enhancement towards the main peak. The peak flux consequently increases by a larger fraction than the fluence, reducing the effective width \(W_{\mathrm{eff}}=A/F_{\mathrm{max}}\). By contrast, the interleaved near-side flares remain direct-image dominated, with little change in their profiles or effective widths.}
    \label{fig:evolution_keplerian_Lightcurve}
\end{figure*}

\appendix
\section{Forward Ray Tracing and Magnification Calculation}
\label{app:kerr_raytracing}

This appendix summarises the mathematical framework used to obtain the image positions, propagation times, and magnifications. We adopt the Kerr geodesic and perturbative mapping formalism of \citet{Forward_RTracing_2025}. For the low-order images considered here, we solve the imaging conditions directly in the \((\lambda,q)\) plane, as illustrated below.

\subsection{Geodesic Quantities and Observer's Image Plane}
\label{appsubsec:kerr_geodesics}
\label{appsubsec:image_plane_order}

We use Boyer-Lindquist coordinates \(x^\mu=(t,r,\theta,\phi)\), with all lengths and times expressed in units of the black hole mass \(M\), as in the main text. Accordingly, we set \(G=c=M=1\) in the equations below, and \(a\) denotes the dimensionless spin \(J/M^2\). The Kerr functions are
\begin{equation}
    \Sigma=r^2+a^2\cos^2\theta,
    \qquad
    \Delta=r^2-2r+a^2.
    \label{eq:app_kerr_functions}
\end{equation}
A photon has conserved energy \(E=-p_t\), axial angular momentum \(L=p_\phi\), and Carter constant \(Q\). In these units, we define
\begin{equation}
    \lambda=\frac{L}{E},
    \qquad
    \eta=\frac{Q}{E^2},
    \qquad
    q=\sqrt{\eta}.
    \label{eq:app_rescaled_constants}
\end{equation}
Together with the initial momentum signs \(b=(\nu_r^s,\nu_\theta^s)\), these quantities specify a null geodesic. Here \(\nu_r=\operatorname{sgn}(p^r)\) and \(\nu_\theta=\operatorname{sgn}(p^\theta)\). The radial and polar potentials are
\begin{subequations}
\label{eq:app_kerr_potentials}
\begin{align}
    \mathcal R(r)
    &=(r^2+a^2-a\lambda)^2
      -\Delta[\eta+(\lambda-a)^2],
    \\
    \Theta(\theta)
    &=\eta+a^2\cos^2\theta-\lambda^2\cot^2\theta.
\end{align}
\end{subequations}
Their zeros determine the turning points, at which the corresponding momentum sign reverses. Since the emission points lie in the equatorial plane, \(\Theta(\pi/2)=\eta\geq0\). We consider ordinary polar motion with \(\eta>0\); exactly equatorial rays are obtained in the limit \(\eta\to0^+\).

At a fixed emission time, we suppress the event index \(i\) and write the source position as \(x_s^\mu=(t_e,r_s,\theta_s,\phi_s)\), where \(t_e=t_i+\tilde t\) follows Equation~\eqref{eq:local_emission_time}. At the observer radius \(r_f=r_o\), the separated geodesic relations are
\begin{subequations}
\label{eq:app_kerr_integral_form}
\begin{align}
    I_r&=G_\theta\equiv\zeta_o,
    \\
    \phi_f-\phi_s&=I_\phi+\lambda G_\phi,
    \\
    \tau_\ell\equiv t_f-t_e&=I_t+a^2G_t.
    \label{eq:app_photon_travel_time}
\end{align}
\end{subequations}
Here \(\zeta\) is the energy-rescaled Mino parameter,
\begin{equation}
    d\zeta=\frac{E}{\Sigma}\,d\sigma,
    \label{eq:app_mino_time}
\end{equation}
where \(\sigma\) is an affine parameter and we choose \(\zeta=0\) at emission. The Mino parameter is distinct from the coordinate propagation time \(\tau_\ell\) of image \(\ell\). The radial integrals are
\begin{subequations}
\label{eq:app_IG_integrals}
\begin{align}
    I_r
    &=\fint_{r_s}^{r_o}
      \frac{\nu_r\,dr}{\sqrt{\mathcal R(r)}},
    \label{eq:app_Ir_def}
    \\
    I_\phi
    &=\fint_{r_s}^{r_o}
      \frac{a(2r-a\lambda)\nu_r\,dr}
           {\Delta\sqrt{\mathcal R(r)}},
    \label{eq:app_Iphi_def}
    \\
    I_t
    &=\fint_{r_s}^{r_o}
      \frac{[r^2\Delta+2r(r^2+a^2-a\lambda)]\nu_r\,dr}
           {\Delta\sqrt{\mathcal R(r)}}.
    \label{eq:app_It_def}
\end{align}
\end{subequations}
The angular integrals are
\begin{equation}
\begin{split}
    G_j&=\fint_{\theta_s}^{\theta_f}
    \frac{w_j(\theta)\nu_\theta\,d\theta}{\sqrt{\Theta(\theta)}},
    \\
    (w_\theta,w_\phi,w_t)&=(1,\csc^2\theta,\cos^2\theta).
\end{split}
\label{eq:app_angular_integrals}
\end{equation}
The slashed integrals follow the actual ray, including all radial and polar turning points; along each segment, \(\nu_r\,dr>0\) and \(\nu_\theta\,d\theta>0\). These integrals admit analytic expressions in terms of elliptic integrals \citep{Lensing_by_KerrBH_2020,Null_Geodesic_KerrExterior_2020}.

For a distant observer at \((r_o,\theta_o,\phi_o)\), the image-plane coordinates are
\begin{subequations}
\label{eq:app_image_plane_alpha_beta}
\begin{align}
    \alpha&=-\frac{\lambda}{\sin\theta_o},
    \\
    \beta&=\nu_\theta^o
    \sqrt{q^2+a^2\cos^2\theta_o-\lambda^2\cot^2\theta_o}.
\end{align}
\end{subequations}
Here \(\nu_\theta^o=(-1)^m\nu_\theta^s\), where \(m\) is the number of polar turning points along the ray. The screen impact coordinates \(\alpha\) and \(\beta\) are expressed in units of \(M\); their corresponding angular coordinates are \((\alpha/r_o,\beta/r_o)\) in the distant-observer approximation.

Following \citet{Forward_RTracing_2025}, we define the image order by
\begin{equation}
\begin{split}
    n&=\frac{G_\theta}{G_\theta^{\rm half}},
    \qquad N=\lfloor n\rfloor,
    \\
    G_\theta^{\rm half}
    &=\int_{\theta_-}^{\theta_+}
      \frac{d\theta}{\sqrt{\Theta(\theta)}}.
\end{split}
\label{eq:app_image_order_def}
\end{equation}
The limits \(\theta_\pm\) are the polar turning points. The image order \(N\), the turning count \(m\), and the individual-image label \(\ell\) have distinct meanings; multiple ray solutions can share the same \(N\). Consistent with Section~\ref{subsec:forward_raytracing}, the light-curve model retains the \(N=0\) and \(N=1\) image contributions.

\subsection{Imaging Conditions and Low-Order Images}
\label{appsubsec:forward_raytracing}

For a source at \(\mathbf{z}_s=(r_s,\theta_s,\phi_s)^{\mathrm{T}}\), each image on an initial-sign branch \(b\) satisfies
\begin{equation}
\begin{split}
    \theta_f(\lambda,q;b,\mathbf z_s)&=\theta_o,
    \\
    \phi_f(\lambda,q;b,\mathbf z_s)&=\phi_o+2\pi k,
    \qquad k\in\mathbb Z,
\end{split}
\label{eq:app_imaging_condition}
\end{equation}
These are the same imaging conditions adopted by \citet{Forward_RTracing_2025}, whose search is parametrized relative to the critical curve. Here, we focus on the \(N=0\) and \(N=1\) images and locate their solutions directly in the \((\lambda,q)\) plane for the four initial-sign branches \(b=(\nu_r^s,\nu_\theta^s)\). By axisymmetry, we set \(\phi_s=0\), so that \(\phi_o=\phi_{o,\mathrm{rel}}\), as defined in Equation~\eqref{eq:observer_relative_azimuth}. Figure~\ref{fig:lambda_q_rootfinding} illustrates how each image corresponds to an intersection of the polar and azimuthal boundary-condition contours. The resulting conserved quantities determine the image position through Equation~\eqref{eq:app_image_plane_alpha_beta} and its propagation time through Equation~\eqref{eq:app_photon_travel_time}.

\subsection{Perturbative Mapping and Image Magnification}
\label{appsubsec:magnification}

The magnification is calculated using the same perturbative mapping formalism as \citet{Forward_RTracing_2025}. A small displacement of the source is accompanied by a change in the photon constants of motion such that the perturbed ray reaches the same observer. We express this mapping using \(q=\sqrt{\eta}\) and the local source-length coordinates defined below.

Let \(\mathbf{p}=(\lambda,q)^{\mathrm{T}}\) and \(\mathbf{f}=(\theta_f,\phi_f)^{\mathrm{T}}\), with the azimuth treated on a continuous local branch. All derivatives below are evaluated at the image solution, with the observer position and the ray branch held fixed. Define
\begin{equation}
    \mathbf{A}
    =\frac{\partial\mathbf{f}}{\partial\mathbf{p}},
    \quad
    \mathbf{B}
    =\frac{\partial\mathbf{f}}{\partial\mathbf{z}_s}.
    \label{eq:app_endpoint_jacobians}
\end{equation}
Linearizing Equation~\eqref{eq:app_imaging_condition} gives
\begin{equation}
\begin{aligned}
    \mathbf{A}\,\delta\mathbf{p}
    +\mathbf{B}\,\delta\mathbf{z}_s=\mathbf{0}, \quad
    \delta\mathbf{p}
    =-\mathbf{A}^{-1}\mathbf{B}\,\delta\mathbf{z}_s,
\end{aligned}
\label{eq:app_implicit_ray_response}
\end{equation}
provided that the local map is differentiable and \(\mathbf{A}\) is nonsingular. Axisymmetry gives
\begin{equation}
    \frac{\partial\theta_f}{\partial\phi_s}=0,
    \quad
    \frac{\partial\phi_f}{\partial\phi_s}=1.
    \label{eq:app_source_azimuth_derivatives}
\end{equation}
The screen response follows analytically from Equation~\eqref{eq:app_image_plane_alpha_beta}:
\begin{equation}
\begin{aligned}
    \mathbf{C}
    &\equiv
    \frac{\partial(\alpha,\beta)}{\partial(\lambda,q)}=
    \begin{pmatrix}
        -\csc\theta_o & 0\\[6pt]
        -\dfrac{\lambda\cot^2\theta_o}{\beta}
        & \dfrac{q}{\beta}
    \end{pmatrix}.
\end{aligned}
\label{eq:app_screen_jacobian}
\end{equation}
This expression applies on a smooth screen branch with \(\beta\neq0\).

To express the source displacement in local length coordinates, we adopt
\begin{equation}
\begin{aligned}
    \delta\mathbf{y}_s=
    \left(
        \delta r_s,\,
        r_s\delta\theta_s,\,
        r_s\sin\theta_s\,\delta\phi_s
    \right)^{\mathrm{T}},\quad
    \delta\mathbf{z}_s
    =\mathbf{S}\,\delta\mathbf{y}_s,
\end{aligned}
\label{eq:app_source_length_coordinates}
\end{equation}
where
\begin{equation}
    \mathbf{S}
    =
    \operatorname{diag}
    \left(
        1,\,
        \frac{1}{r_s},\,
        \frac{1}{r_s\sin\theta_s}
    \right).
    \label{eq:app_source_coordinate_scaling}
\end{equation}
These coordinates specify a local Euclidean-length convention for the source size, without a conversion to proper lengths in the emitter's comoving tetrad. Combining the source and screen responses gives the dimensionless \(2\times3\) mapping matrix
\begin{equation}
\begin{aligned}
    \mathbf{M}_\ell
    =-\mathbf{C}\mathbf{A}^{-1}\mathbf{B}\mathbf{S},\quad
    (\delta\alpha,\delta\beta)^{\mathrm{T}}
    =\mathbf{M}_\ell\,\delta\mathbf{y}_s.
\end{aligned}
\label{eq:app_source_to_image_mapping}
\end{equation}
The factor \(\mathbf{S}\) ensures that radial and angular perturbations are compared on the same source-length scale.

Consider an infinitesimal spherical source of radius \(R_{\mathrm{src}}\) in these local length coordinates. The linear map projects it to an ellipse with semiaxes \(R_{\mathrm{src}}s_{1,\ell}\) and \(R_{\mathrm{src}}s_{2,\ell}\), where \(s_{1,\ell}\) and \(s_{2,\ell}\) are the two singular values of \(\mathbf{M}_\ell\). Relative to the reference projected area \(\pi R_{\mathrm{src}}^2\), the positive geometric magnification is
\begin{equation}
    \mu_\ell
    =
    \frac{\pi R_{\mathrm{src}}^2s_{1,\ell}s_{2,\ell}}
         {\pi R_{\mathrm{src}}^2}
    =s_{1,\ell}s_{2,\ell}
    =
    \sqrt{\det\!
    \left(
        \mathbf{M}_\ell\mathbf{M}_\ell^{\mathrm{T}}
    \right)}.
\label{eq:app_area_magnification}
\end{equation}
The source radius cancels, giving the local point-source magnification. Finite-source averaging is not included.

For event \(i\), this calculation is repeated along the source trajectory in Equation~\eqref{eq:source_trajectory_local_time} at each emission-time offset \(\tilde t=t_e-t_i\). The resulting propagation time \(\tau_{i\ell}(\tilde t)\) and magnification \(\mu_{i\ell}(\tilde t)\) enter the arrival-time mapping and image-flux prescription in Equations~\eqref{eq:image_observer_time_mapping} and \eqref{eq:image_flux_parametric}, respectively. As in the main text, the geometric magnification is used as a flux weight for uniform projected source brightness, with Doppler and gravitational frequency-shift transfer factors neglected.

\section{Angular Magnification Profiles and Flare Envelopes}
\label{appsubsec:magnification_envelopes}

The one-dimensional cuts in Figure~\ref{fig:envelope_profile_comparison} illustrate how the total magnification varies with observer-relative azimuth. At \(\theta_o=85^\circ\), increasing \(r_s\) lowers the peak magnification and narrows, or removes, the interval above a fixed magnification threshold. At fixed \(r_s=50\,M\), moving the observer closer to the equatorial plane raises the peak and widens the above-threshold interval. As Lense-Thirring precession shifts successive disk crossings in azimuth, each branch samples a profile of this form, producing the slowly varying amplitude envelope in Figure~\ref{fig:Static_LightCurve}. The full light-curve calculation also includes the small variations in crossing radius caused by apsidal precession.

The width above a fixed threshold follows from the angular scaling \(\delta_{\mathrm{th}}\propto r_s^{-1/2}\) discussed in Section~\ref{subsec:lensing_possibility}. Approximating the threshold contours as locally circular, the azimuthal half-width at a fixed inclination is
\begin{equation}
    \delta\phi_{\mu_{\mathrm{th}}}(r_s,\theta_o)
    \simeq
    \Biggl[
    \delta_{\mathrm{th}}^2(50\,M)
    \left(\frac{r_s}{50\,M}\right)^{-1}
    -
    \left(\theta_o-\frac{\pi}{2}\right)^2
    \Biggr]_+^{1/2},
    \label{eq:lensing_cut_half_width}
\end{equation}
where \(\delta_{\mathrm{th}}(50\,M)\) is the angular radius of the threshold contour at \(r_s=50\,M\), \([x]_+=\max(x,0)\), and all angles are in radians. The approximate full width above the threshold is \(2\delta\phi_{\mu_{\mathrm{th}}}\). The arrows in Figure~\ref{fig:envelope_profile_comparison} mark the measured one-sided widths for \(\mu_{\mathrm{th}}=3.0\), from the actual profile maximum to the right-hand threshold crossing. Measuring from the actual maximum allows for its small spin-induced displacement from \(\phi_{o,\mathrm{rel}}=180^\circ\).

\section{Evolution of Successive Flare Profiles}
\label{subsec:keplerian_evolution}

In the Keplerian-hotspot model, Lense-Thirring precession changes the disk-crossing azimuth and hence the initial hotspot phase. As the hotspot moves during each flare, successive events on the same branch traverse different segments of the lensing pattern. The time-dependent magnification therefore shifts relative to the intrinsic emission, changing the individual flare profiles.

Figure~\ref{fig:evolution_keplerian_Lightcurve} shows two consecutive segments of the resulting flare train. In the interval shown, odd events are far-side flares and even events are near-side flares. Along the far-side sequence, the magnification maximum occurs progressively earlier relative to the intrinsic luminosity maximum. The associated flux excess shifts from the declining wing towards the main peak, bringing the two peaks closer together. The near-side flares remain direct-image dominated and retain nearly the same profile.

\bibliographystyle{apsrev4-1}
\bibliography{refs}
\end{document}